\documentclass[hyper]{JHEP}
\usepackage{epsf}
\usepackage{amsmath}
\usepackage{amssymb}
\usepackage{amsfonts}

\newcommand{\tr}{{\rm tr}}
\newcommand{\CO}{{\cal O}}
\newcommand{\notsubset}{\subset\kern-0.9em /\kern+0.3em}
\newcommand{\state}[1]{|#1\rangle}
\newcommand{\ii}{{\it i}}
\newcommand{\Ann}{\hbox{Ann}}
\newcommand{\qcomm}[2]{\left[#1,#2\right]_q}

\DeclareFixedFont{\xiiss}{OT1}{cmss}{m}{n}{12}
\DeclareFixedFont{\ixss}{OT1}{cmss}{m}{n}{9}
\DeclareFixedFont{\cmrnine}{OT1}{cmr}{m}{n}{9}
\newcommand{\field}[1]{\mathbb{#1}}
\newcommand{\BC}{{\field C}}
\newcommand{\BR}{{\field R}}
\newcommand{\BZ}{{\field Z}}

\newcommand{\CCs}{\hbox{\ixss C\kern-.4emI}}
\newcommand{\ZZs}{\hbox{\ixss Z\kern-.4emZ}}

\newcommand{\CA}{{\cal A}}
\newcommand{\CB}{{\cal B}}
\newcommand{\CI}{{\cal I}}
\newcommand{\CM}{{\cal M}}
\newcommand{\CZ}{{\cal Z}}
\newcommand{\CS}{{\cal S}}
\newcommand{\CR}{{\cal R}}
\newcommand{\ZA}{\CZ\CA}

\newcommand{\littlefig}[2]{
	\epsfxsize=#2in
	\epsfbox{#1}
}
\preprint{ILL-(TH)-00-04}
\title{Marginal and Relevant 
Deformations of N=4 Field Theories and Non-Commutative 
Moduli Spaces of Vacua
}

\author{David Berenstein,\thanks{\email{berenste@pobox.hep.uiuc.edu}}
Vishnu Jejjala,\thanks{\email{vishnu@pobox.hep.uiuc.edu}} 
and Robert G. Leigh\thanks{\email{rgleigh@uiuc.edu}} 
\\
Department of Physics\\
University of Illinois at Urbana-Champaign\\
Urbana, IL 61801}

\abstract{
We study marginal and relevant supersymmetric deformations of the $N=4$
super-Yang-Mills theory in four dimensions. Our primary innovation is
the interpretation of the moduli spaces of vacua of these theories as
non-commutative spaces. The construction of these spaces relies on the
representation theory of the related quantum algebras, which are
obtained from $F$-term constraints.

These field theories are dual to superstring theories propagating on
deformations of the $AdS_5\times S^5$ geometry. We study $D$-branes
propagating in these vacua and introduce the appropriate notion of
algebraic geometry for non-commutative spaces. The resulting moduli
spaces of $D$-branes have several novel features. In particular, they
may be interpreted as symmetric products of non-commutative spaces. We
show how mirror symmetry between these deformed geometries and orbifold
theories follows from T-duality.

Many features of the dual closed string theory may be identified within
the non-commutative algebra. In particular, we make progress towards
understanding the K-theory necessary for backgrounds where the
Neveu-Schwarz antisymmetric tensor of the string is turned on, and we
shed light on some aspects of discrete anomalies based on the
non-commutative geometry.
}


\begin{document}

\section{Introduction}\label{sec:intro}

The study of the deformations of the $N=4$ $U(M)$ supersymmetric theory
in four dimensions is of interest from several points of view. This
theory is superconformally invariant, and it has been known for some
time that exactly marginal deformations of this theory exist (Ref. \cite{LS}
and references therein)
and should be described by interacting superconformal field theories
(CFT). These CFT's are largely unexplored.

In the large $M$-limit, the deformations of the $N=4$ theory have a nice
description in terms of a supergravity dual\cite{M,W,GKP} and are
obtained by the addition of operators which modify the boundary
conditions at infinity. Each of the renormalizable deformations are
reflected in the $AdS/CFT$ correspondence through backgrounds for
massless and tachyonic excitations, including both $RR$ and $NS$
fields\cite{KRN}.

Among the marginal deformations, of particular interest is the
$q$-deformation
\begin{equation}
W_q=\tr\ \left( \phi_1\phi_2\phi_3-q\phi_2\phi_1\phi_3\right)
\end{equation}
which is a deformation of the superpotential by the symmetric invariant
preserving $N=1$ supersymmetry and a $U(1)^3$ global symmetry. For
special values of $q$ these theories are described by the near-horizon
geometries of orbifolds with discrete torsion\cite{D,BL}. It was
conjectured in Ref. \cite{BL} that these orbifold theories are related
by mirror symmetry to string theories on $S^5$-deformations of
$AdS_5\times S^5$.

There are also relevant deformations which carry the theory away from
the ultraviolet fixed point CFT. In some cases, the infrared theory is
of interest---a prime example being the deformation by rank-one mass
terms\cite{LS,AILS}. From the supergravity point of view, the
renormalization group flow is encoded as a dependence of the background
on the radial scaling variable\cite{FW,GPPZ}.

With a rank-three mass matrix, the field theory has been analyzed in
many papers\cite{VW2,DW,S1,Dorey,DK}. More recently\cite{PS}, the
supergravity duals of these theories have been analyzed. There it was
noticed that 5-brane sources resolve the would-be singularity in the
dual supergravity background, an application of the dielectric
effect\cite{RM}.

In this paper, we will begin an exploration of these field theories
obtained by marginal and relevant deformations of the $N=4$ theory. The
analysis will concentrate on the classical vacua, particularly those
aspects which depend upon holomorphic quantities. We introduce a new way
of thinking about these moduli spaces that should be of quite general
applicability.

Normally, the vacua of a supersymmetric gauge theory are parameterized
using gauge invariant holomorphic polynomials in the fields. This is
attractive because of the gauge invariance, but it is also unwieldy. As
$M$ increases, the number of independent invariants increases
dramatically. The $F$-term constraints on vacua are given, on the other
hand, directly in terms of holomorphic matrix equations, and the
proposal centres around using this description directly. Matrix
variables have a number of technical advantages, principally that the
analysis is independent of their dimension, $M$. The main problem with
this approach is gauge invariance---the $D$-term constraints must be
applied separately.

Given this, the $F$-terms can be thought of as a set of constraints on
the algebra of $M\times M$ matrices. Generally, this is a
non-commutative algebra. There is a technical simplicity to the choice
of renormalizable superpotentials, namely that the constraints are
quadratic, and this simplifies the algebraic analysis significantly. The
constrained algebras that appear here in some cases bear some
resemblance to algebras considered in the literature on quantum groups
(see for example, Refs. \cite{Manin,BigQBook}).

A related problem is the behavior of $D$-branes in dual descriptions of
these field theories. For small deformations, these duals are close to
$AdS_5\times S^5$, and therefore the moduli space of $D$-branes also has a
description in this framework. The moduli space of probe $D$-branes is
roughly a symmetric product space; indeed, classically, we can think of
a single $D$-brane as moving on the moduli space of the corresponding
field theory, and the moduli space for multiple branes can often be
related to the direct product of this space, modded out by the
permutation group. Realizing all aspects of the field theory analysis in
these dual descriptions gives insight into many non-trivial aspects of
$D$-brane geometry. In particular, a clear understanding of these issues
reveals a T-duality transformation which realizes mirror
symmetry\cite{BL} between near-horizon geometries and orbifold theories.

In the present context, these remarks lead to the notions of
non-commutative moduli spaces of vacua, and moduli spaces of $D$-brane
configurations are symmetric products of a non-commutative space. We
construct these notions algebraically; for example, points in the
non-commutative space correspond to irreducible representations of the
algebra or equivalently to maximal ideals with special properties. As we
show later, these notions have some tremendous advantages over the
standard points of view. In particular, it is often the case that we can
think of a commutative subring (built out of the center of the algebra)
as a sort of coarse view of the full moduli space. In fact, the
phenomenon of $D$-brane fractionation at singularities follows precisely
this rule: the fractionation is present in a commutative description,
but from the full non-commutative point of view, the fractional nature
is more readily understandable.

It is clear that from a string theory point of view, it is the open
strings that see this non-commutative structure directly\cite{CDS}.
Closed strings appear naturally within this framework as single trace
operators\cite{W}, and thus should see only the commutative part of the
space\cite{SW}. The remnant of non-commutativity in the closed string
sector is the presence of twisted states.

In general, when one studies more general configurations of $D$-branes,
they should correspond to algebraic geometric objects and classes in
K-theory\cite{MM,W3,MW}. Because of our emphasis, one needs to develop a
non-commutative version of algebraic geometry and K-theory. K-theory in
this context is provided by the algebraic K-theory of the
non-commutative ring. We give the rudimentary structure of such a
definition of algebraic geometry; this definition apparently differs
from others given in the mathematics literature\cite{AZ,Rosenberg,Kap},
but we believe our proposal is more natural, as dictated by string
theory.

Our version of non-commutative geometry is clearly different than that
which has been recently studied extensively (see for example
\cite{CDS,HI,MR,SW,RS,MRS} and citations thereof). In that case, the
non-commutativity occurs in the base space of a super-Yang Mills theory,
whereas here it is in the moduli space, namely the directions transverse
to a brane. In many cases, the boundary state formalism (for a review,
see Ref. \cite{Gab}) is convenient to describe $D$-branes, but we do not
use that technology here. It would be interesting to generalize our
discussion to that formalism, although it is not clear to us how to
solve for the boundary states in the absence of a spectrum-generating
algebra.

The paper is organized as follows. In Section \ref{sec:problem}, we
review the structure of marginal and relevant deformations of the $N=4$
theory, and discuss the interpretation of supersymmetric vacua in terms
of non-commutative geometry. (We focus throughout on $U(M)$ gauge
groups.) We also review the map between the superpotential deformations
and supergravity backgrounds. In Section \ref{sec:algebra}, we give our
construction of non-commutative algebraic geometry and the resulting
K-theory. This section is mathematically intensive; in order that the
casual reader may skip this section if desired, we provide at the
beginning, an overview of the key structures. In Section
\ref{sec:discrete}, we investigate the vacua of various field theories,
using the non-commutative formalism. We begin with the $q$-deformed
theory, and then investigate this theory with (a) a single mass term,
(b) a mass term and a linear term, (c) three arbitrary linear terms, and
(d) three mass terms. In each case, we work out the representation
theory. We also consider the general case, and in particular consider
the effects of the other independent marginal deformation. The general
case is quite difficult, but we are able to identify a few interesting
properties.

In Section \ref{sec:Dbranes}, we turn our attention to string theory.
For the $q$-deformed theory, the field theory predicts new branches in
moduli space for arbitrarily small values of $q-1$. To realize this
branch in string theory, we need to consider BPS states corresponding to
$D5$-branes with 3-brane charge in the deformed backgrounds; the physics
here is reminiscent of the dielectric effect\cite{RM}, but is more
general. The new branch of moduli space is identified as a $D5$-brane
wrapped on a degenerating 2-torus. One obtains a natural 2-torus
fibration of the 5-sphere; T-duality on this torus leads to the mirror
orbifold theory.

In Section \ref{sec:closedstrings}, we consider the problem of
identifying closed string physics directly from the field theory
description. Closed string states are naturally identified with single
trace operators.
In this section, we also note several features of interest, including
connections to quantum groups and the K-theory of the non-commutative
geometry.

In Section \ref{sec:conclus}, we make some final remarks and indicate
avenues for further research.

\section{Field theory deformations}\label{sec:problem}

Our first objective will be to analyze the marginal and relevant
deformations of the $N=4$ super Yang-Mills (SYM) theory in four
dimensions, with gauge group $U(M)$. As usual, we write this in terms of
an $N=1$ SYM theory with three adjoint chiral superfields $\phi_i$,
$i=1,2,3$, coupled through the superpotential
\begin{equation}
W = g\ \tr\left( \left[\phi_1,\phi_2\right]\phi_3\right).
\end{equation}

If we choose to preserve $N=1$ SCFT, there is a moduli space of marginal
deformations\cite{LS}, given by a general superpotential of the form
\begin{equation}\label{eq:genmarg}
W_{marg}=a\ \tr\left( \phi_1\phi_2\phi_3-q\phi_2\phi_1\phi_3+\frac{\lambda}{3}
\left(\phi_1^3+\phi_2^3+\phi_3^3\right)\right).
\end{equation}
The Yang-Mills coupling $g$ measures how strongly interacting the
theory is and is a function of $a,q,\lambda$ such that
each of the $\beta$-functions are zero. The structure of the moduli
space of vacua depends only on $q,\lambda$. 

We will also consider relevant deformations of the form
\begin{equation}\label{eq:genrelev}
W_{rel}=c_1 \tr(\phi_1^2) + c_2\tr(\phi_2^2+\phi_3^2)
+\sum_j\zeta_j\tr(\phi_j).
\end{equation}
For $q\neq 1$, general quadratic polynomials may always be brought
to this form after a change of variables.

The vacua of the theory are found by solving the $F$-term constraints
\begin{equation}
\frac{\partial W}{\partial \phi_{j}} = 0.
\end{equation}
In the present cases, these are quadratic matrix polynomial equations in 
the $\phi_j$
\begin{eqnarray}\label{eq:Fterms}
\phi_1\phi_2-q \phi_2\phi_1&=& -\lambda \phi_3^2 - 2 c_2 \phi_3-\zeta_3\\
\phi_2\phi_3-q \phi_3\phi_2 &=& - \lambda \phi_1^2 - 2 c_1 \phi_1 -\zeta_1\\
\phi_3\phi_1-q \phi_1\phi_3 &=& - \lambda \phi_2^2 - 2 c_2 \phi_2 -\zeta_2\label{eq:Fterms3}
\end{eqnarray}
These matrix equations are independent of $M$. In general, solutions will 
consist of a collection of points, but at special values of parameters, we
get a full moduli space of vacua.

The equations (\ref{eq:Fterms})--(\ref{eq:Fterms3}) are a quite general
class of relations. Note in particular that when $q=1$, we have a
Poisson bracket structure, whereas if in addition $\lambda =\zeta_i =0$,
we find $SU(2)$ commutation relations. When $q\neq 1$ and/or
$\lambda\neq 0$, with $c_j=\zeta_j=0$, the algebra is that of a quantum
plane\cite{Manin}. When $q\neq1$ and $\lambda=c_j=0$, these are
$q$-deformations of Heisenberg algebras.

The moduli space of vacua is usually parameterized in terms of
gauge-invariant polynomials in the fields $\phi_j$. This has the feature
that the non-holomorphic $D$-term constraints are automatically
satisfied. The down side is that for large $M$, the number of
polynomials required becomes very large, and when perturbations are
present, the description of the space becomes quite complicated.
Instead, we will choose to describe the moduli space of vacua directly
in terms of matrix variables. This has the virtue that the equations are
independent of $M$, as noted. Thus instead of considering the moduli
space of vacua as an algebraic variety, for general values of
parameters, we should think of this as a non-commutative algebraic
variety.

Understanding the vacua of these theories then is equivalent to
understanding the non-commutative geometry defined by the relations
(\ref{eq:Fterms})--(\ref{eq:Fterms3}). The $\phi_j$ can be thought of as
the generators of the corresponding non-commutative algebra. $M\times M$
matrices which satisfy the relations are an $M$-dimensional
representation of the abstract algebra. The general problem at hand then
is to study the representation theory of the algebra. The basic
representations of interest are those that are irreducible; given a
finite set of such solutions $(\phi_1^i, \phi_2^i, \phi_3^i)$ labeled by
$i$, then
\begin{equation}\label{eq:thiseq}
\tilde \phi_k = \oplus_i\ \phi_k^i
\end{equation}
is also a solution of the matrix equations.

It is important to keep in mind however that we must also consider the
$D$-term constraints. It is well-known\cite{LT} that for every solution
of the $F$-term constraints, there is a solution to the $D$-terms in the
completion of the orbit of the complexified gauge group $SL(M,\BC)$. If
the solution occurs at a finite point in the orbit, then we get a true
vacuum. If it occurs in the completion of the orbit (at infinity in the
complexified gauge group), we need to check that the solution does not
run away to infinity.

\subsection{Relation to Deformations in $AdS_5\times S^5$ Geometry}

Since the $N=4$ theory is related to superstring theory on $AdS_5\times
S^5$, the marginal deformations correspond to deformations of $S^5$. In
particular, these are related to massless states\cite{KRN,W} in the
5-dimensional supergravity. They transform in the ${\bf 45}$ of $SU(4)$
R-symmetry and are related to vevs for harmonics of $RR$ and $NSNS$ fields,
$F^{RR}_{(3)}$ and $H^{NS}_{(3)}$, along the 5-sphere. Similarly, 
the relevant deformations correspond to tachyonic excitations of the
5-dimensional supergravity, and transform in the ${\bf 10}$ of $SU(4)$.

When $q$ is a root of unity, we will often for convenience say that $q$ is
rational. In this case, the $q$-deformation is known to be dual to
the near-horizon geometry of $D$-branes on an orbifold with 
discrete torsion, $\BC^3/(\BZ_n\times\BZ_n)$. 

The moduli space of vacua of the field theory is the moduli space of
$D$-branes. Because of the $RR$ and $NSNS$ backgrounds, 
$D$-branes move on a non-commutative space\cite{CDS}. For small enough
deformations, we expect the $AdS_5\times S^5$ geometry to be a close
approximation and we can interpret the eigenvalues of matrices as
the positions of $D$-branes, \`a la matrix theory\cite{BFSS}.

Note that the superpotentials that we are considering are single trace
operators. This suggests that these operators correspond
to effects that may be seen in classical supergravity. 
This may be understood by looking at how background couplings
to $D$-branes behave at weak coupling. The leading effect comes from
a disk diagram as shown in Figure \ref{fig:tadpole} where $V$ is
the background vertex.
\FIGURE[1]{
\littlefig{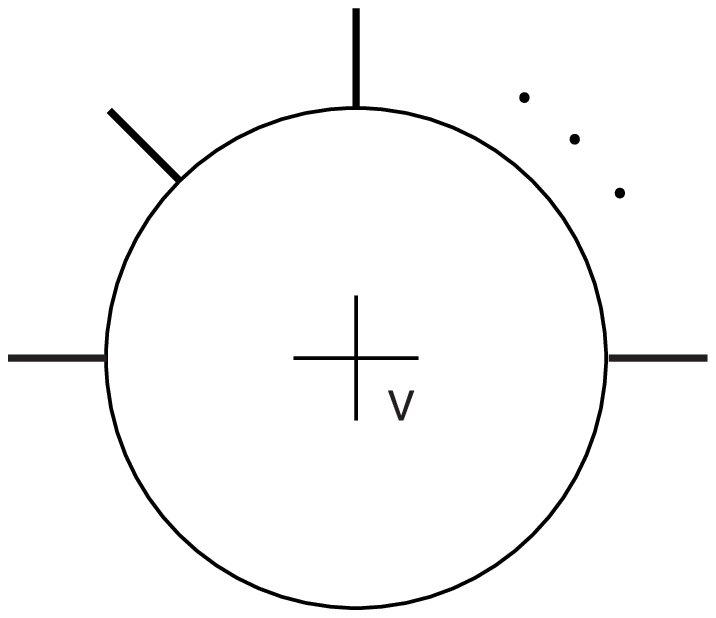}{1.5}
\caption{Tadpole calculation of superpotential}\label{fig:tadpole}}
Multiple trace operators would then correspond to string loop diagrams,
and are therefore suppressed by powers of $g_{str}$.\footnote{We assume 
 a large but finite number of branes, so that
relations between traces of finite matrices appear only for very
irrelevant perturbations.} It is not clear that the string generates
these effects perturbatively, but to avoid them we could work at
weak string coupling. It is also possible that a non-perturbative
non-renormalization theorem might keep 
multiple trace operators equal to zero, at least up to some number
of derivatives.

\vspace{1cm}

\section{Non-Commutative Algebraic Geometry}\label{sec:algebra}

\subsection{Overview}\label{sec:overv}

As discussed above, we usually think of the moduli spaces of vacua as
varieties, namely commutative algebraic geometric objects. Because the
$F$- and $D$-term constraints may be recast in matrix form, it is more
convenient to think of the same space as a non-commutative object.
Although the physical problem to solve is the same and the space of
solutions is the same, the non-commutative interpretation invokes extra
structure (namely, the commutative algebra of individual matrix elements 
is organized into the non-commutative algebra of matrices). 
Because $D$-brane solutions are associated with algebraic
geometric objects in general\cite{HM} (and see \cite{Drev} for a review), 
we need a formulation of non-commutative
algebraic geometry which captures $D$-brane physics correctly.

Several different versions of non-commutative algebraic geometry have
been discussed in the mathematics literature\cite{AZ, Rosenberg, Kap},
but none of these seem natural in the present context. In this section,
our aim is to describe a definition of algebraic geometry appropriate to
the moduli spaces of vacua of $D$-branes in the field theory limit.

We want to understand the holomorphic structure of the moduli space, so
we will only concentrate on the $F$-term equations and will assume that
given a solution to the $F$-terms, there is a solution to the $D$-term
equations. From the physics point of view, we confine ourselves to those
properties which are protected by supersymmetry; from a mathematical
viewpoint, these are holomorphic structures and can be described in
terms of algebraic geometry. Ordinarily, non-commutative geometries are
related to $C^*$-algebras\cite{Connes}, which include the adjoint
operation; this is not a natural operation in a holomorphic framework,
and we will discard it for the considerations of this paper. In
supersymmetric theories, the holomorphic and anti-holomorphic features
couple only through $D$-terms, and these effects are in general not
protected by supersymmetry. In discarding the $D$-terms, we lose
information about the metric in moduli space, but not topological
features. Thus we need a framework where we can do non-commutative
algebraic geometry without $C^*$-algebras.

In the rest of this subsection, we give a brief
outline of the mathematics involved, and its physical interpretations.
For the reader who wishes to skip the details of the mathematics, 
this overview should suffice, and one may proceed to
Section \ref{sec:discrete}. 

The building blocks for solutions are the finite dimensional irreducible
representations of the algebra, as in eq. (\ref{eq:thiseq}). In the
non-commutative algebraic geometry that we will describe, these are
defined to be points\cite{Connes,Landi}. In matrix theory, 
non-commuting matrices are interpreted as extended objects; here, these
are considered point-like objects. In orbifold theories, D-branes in the bulk
are also considered to be point-like even though they can be built out of
fractional branes. 

In our construction the center (the commutative
sub-algebra of the Casimir operators) plays a pivotal role. In
particular, one expects\cite{SW} that for a given background in string
theory, there are two descriptions, a commutative version relevant to
closed strings and a non-commutative version for open strings. In our
best-understood examples, the commutative space is the algebraic
geometry associated to the center. By Schur's lemma, on every
irreducible representation the Casimirs are proportional to the
identity. Because of this, one finds a map between non-commutative
points and commutative points. For ``good'' algebras, the
non-commutative geometry covers the commutative geometry; in this case,
we will say that the non-commutative algebra is {\it semi-classical}. In
this case we can think of the commutative space as a coarse-grained
version of the full non-commutative space. For semi-classical algebras,
our interpretation of irreducible representations as points is equivalent
to the point-like properties of $D$-branes in orbifolds. In other cases, the 
non-commutative geometry may have little relation to the commutative
geometry and it is not clear that one should interpret the $D$-brane
states as pointlike.

A general solution of the $F$-term constraints is a direct sum of
irreducible representations, and thus the natural non-commutative
structure is an unordered finite collection of points. For commutative
algebras, we would interpret this as a symmetric product space, and we
will carry over this name in the non-commutative case. 
This symmetric product structure leads directly to an interpretation
of $D$-brane fractionation at a singularity, whereby an irreducible 
representation can be continuously deformed and becomes reducible
at a certain point. In this sense, in the non-commutative version,
single-particle and multi-particle states are continuously connected.
In commutative geometry on the other hand, this process would be singular.

The remainder of the formal discussion deals
with an extension of this construction to subvarieties, sheaves,
and the algebraic K-theory of the ring, relevant to an understanding
of extended $D$-branes. The discussion presented here lays the outline
for non-commutative algebraic geometry; a full account
will appear elsewhere\cite{BJL}.

\subsection{Preliminaries: Points and Topology}

Consider an associative algebra $\CA$ over the complex numbers $\BC$,
generated by a set of operators subject to some relations. (In all the
examples we consider, we will have three generators and a set of
quadratic relations.) As is standard, the non-commutative algebra should
be thought of as a ring of functions on some affine non-commutative
space\cite{Connes}. Ring homomorphisms should correspond to holomorphic
maps between affine non-commutative geometries. We will assume that all
rings are Noetherian (so that any ideal always has a finite basis) and
that the algebra is polynomial.

Given the matrix equations (\ref{eq:Fterms})--(\ref{eq:Fterms3}), we
want to find solutions in terms of $M\times M$ matrices (with
unspecified $M$). That is, we are interested in representations of the
algebra, and we will assign a geometrical space to these solutions.

An element  $a\in\CA$ is central if it commutes with every other element
in $\CA$; that is, $a$ is a Casimir of the algebra. We usually think of
the Casimir operators as sufficient to define a representation ({\it
e.g.}, as in the finite dimensional representations of $SL(2,\BC)$) and
thus will pay particular attention to the center of the algebra, denoted
$\CZ\CA$.

Since $\CZ\CA$ is commutative, we can associate an ordinary commutative
space to it, which is the general philosophy behind algebraic geometry
(see for example, Ref. \cite{Hart}). We interpret the center of the
algebra as a coarse description of the full non-commutative geometry.
The picture we have is that there is a map between the non-commutative
geometry to the commutative one which forgets some of the structure
(namely the functions that don't commute). The natural inclusion
$\ii:\CZ\CA \to \CA$ is to be thought of as the pullback of functions
from the commutative space to the non-commutative space.

To describe the non-commutative space, we need to define the notions of
points and open sets and to impose a topology. Loosely speaking, a point
will be a solution of the constraint equations in finite matrices (just
as points in varieties are solutions of the equations defining the
variety). In commutative algebra, points are interpreted as maximal
ideals of the algebra, and we want to incorporate both of these notions
in our definition of a point.

Let us now be precise. A representation $R$ of dimension $M$ of the
algebra $\CA$ is an algebra homomorphism $\mu$ from $\CA$ to the algebra
of $M\times M$ matrices  ({\it i.e.}, $\mu$ respects the addition,
product, and multiplication by scalars). For the map to be well defined,
the relations ($F$-term constraints) must be satisfied in terms of the 
$M\times M$ matrices.

A representation is irreducible if there is no linear subspace of
$\BC^M$ which is invariant under multiplication by all the elements of
the image of the algebra $\mu(\CA)$. The representation is reducible
otherwise. If a representation is irreducible, then the map is such that
we have an exact sequence
\begin{equation}
\CA\stackrel{\mu}{\to} M_M(\BC)\to 0
\end{equation}
with the map $\mu$ defined by the representation of the algebra. The kernel of
this map is a double-sided ideal $\CI$ of $\CA$ to which the
representation is associated, and we have the isomorphism
\begin{equation}
\CA/\CI \sim M_M(\BC).
\end{equation}
This isomorphism is non-canonical (two representations are identified
if they lie in the same orbit of the group $GL(M,\BC)$ by similarity
transformations).

The space associated to an algebra is constructed from the irreducible
representations of $\CA$ as follows. To each irreducible representation
of finite dimension $M$, there is an associated ideal in the algebra
$\CA$, namely the ideal $\CI= \ker(\mu)$. $\CI$ is a double-sided
maximal ideal and is declared to be a point. This definition is borrowed
from \cite{Connes, Landi}, but without the $C^*$-algebra framework. In
general, one would also allow infinite dimensional representations of the
algebra; in that case, we would need some sense of convergence for
sequences. For our physical problem, we are interested only in finite
dimensional representations, and so we will simply discard this
possibility. As a result, we have a space which is better behaved from
an algebraic standpoint. Irreducible representations are considered
equivalent if they are related by a  change of basis ({\em i.e.}, by orbits of
the group $GL(M,\BC)$). This equivalence is the fact that we have in
supersymmetric field theories a complexified gauge group, and each
point has an associated maximal double-sided ideal $\CI$ of the
algebra $\CA$ such that $\CA/\CI$ is non-canonically isomorphic to the
algebra of $M\times M$ matrices. The variety associated to $\CA$ will be
labeled $\CM_\CA$.

A closed set is defined in terms of an arbitrary double-sided ideal
$\CI'$ in $\CA$, as one does to define the Zariski topology of a space.
A closed set is the collection of points (given by maximal ideals $\CI$)
which contain $\CI'$. By definition, points are closed sets, as one
takes the maximal ideal $\CI$ associated to the point.

The union of two closed sets corresponds to the double-sided ideal $\CI
= \CI_1\cap \CI_2$, and the intersection of two closed sets corresponds
to the double-sided ideal $\CI_1+\CI_2$, which is the direct sum of the
ideals. Indeed, direct sums may be extended to an infinite number of
ideals, so arbitrary intersections and finite unions of closed sets are
closed and define a topology on the set of points. This should be
thought of as a model for the  definition of the geometry, and the
construction mimics the construction of algebraic varieties over $\BC$
as much as possible.

Note that the definition of a point has the following technical property. 
A point is Morita-equivalent to a point in a commutative
algebraic variety. This is important for K-theory considerations, which we
return to in a later subsection.

\subsection{Naturalness of Symmetric Spaces}

So far, we have defined a non-commutative space together with some
topology. Given these definitions, additional structure naturally
emerges, as we now discuss.

When one has the ring of functions of a variety, one can pull back
functions between maps of varieties. Thus, a map between non-commutative
spaces will correspond to ring homomorphisms. If we take two rings $\CA$
and $\CB$ and consider a ring homomorphism $\varphi:\CA\to \CB$, it will
correspond to a continuous map from $\CM_\CB\to \CM_\CA$.

Now consider a point $x\in\CM_{\CB}$. By construction, it corresponds to
an irreducible representation of $\CB$ in $M\times M$ matrices for some
$M$. We label the corresponding representation $r_x$, {\it i.e,} a
homomorphism $\mu_x: \CB\to M_M(\BC)$. Thus we have a diagram of maps
\begin{equation}
\CA \stackrel{\varphi}{\to}\CB\stackrel{\mu_x}{\to}M_M(\BC).
\end{equation}
We wish to find the image of the point $x$ in $\CM_\CA$. 
By composing arrows, we get a
natural representation of $\CA$ in the ring of $M\times M$ matrices. The
natural image of $x$ is the kernel of the composition map, but it is
important to note that the corresponding $M\times M$ representation of
$\CA$ might be reducible. In general, to find irreducible
representations associated to a given reducible one, we need to consider
the composition series of the representation $R$.

That is, given a reducible representation $R$, there is an invariant
linear subspace $R'$, and we have an exact sequence of vector
spaces:
\begin{equation}
0\to R'\to R \to R/R' \to 0.
\end{equation}
By construction the dimensions of $R'$ and $R/R'$ are smaller than that
of $R$. If any of $R'$, $R/R'$ is reducible, we repeat the procedure.
Eventually, we obtain a collection of irreducible representations of
$\CA$, and any two such decompositions contain the same irreducibles.

The image of $x$ should be considered a positive sum of points in
$\CM_\CA$ with multiplicities given by the number of times a particular
irreducible representation of $\CB$ appears. Thus the map is
multivalued, given the description used so far. If we wish to make such
maps single-valued, we can either restrict the choice of maps, or modify
the definition of a point. The latter possibility is most natural, and
there is an obvious choice. We should consider instead a new space
consisting of the free sums of points with coefficients in $\BZ^+$.
These sums of points are generated by the points of $\CM$ such that the
sums are finite. We should consider the maps between two such spaces  as
linear transformations between the two formal sums. Notice that the
formal positive sums of $n$ points in a commutative variety $\CM$
corresponds to the  symmetric product $\CM^n/S_n$. Hence the natural
object to understand in this version of non-commutative algebraic
geometry is the symmetric product of the space $\CM_\CA$. This is also
the framework necessary for matrix theory and matrix string
theory\cite{BFSS,DVV}, and this is why we find it a very appealing
aspect of the construction. We will denote this symmetric product space
by $\CS\CM_\CA$. $\CM_\CA$ is a subset of its symmetric space, and it is
the set which generates the formal sums.

There is a grading present here which gives a notion of the degree of a
point, which we denote $\deg(x)$. There is a natural map from the formal
sums of points to $\BZ$; namely, for each irreducible representation
$x\in \CM_\CA$, we consider the map that assigns to $x$ the dimension of
the representation that $x$ is associated with (that is, the character
$\tr_{\mu_x} 1$). This extends by linearity to the symmetric product
space, and the maps between the sums of points are such that they are
degree-preserving. Indeed, given any function on the space (an element
of the ring $a\in\CA$), we consider the invariant of the function $a$ at
the point $x$, $\tr_{\mu_x} a$. The trace is linear, and independent of
the choice of basis for the local matrix ring and can therefore be
extended to direct sums of representations ({\it i.e.,} to the space
$\CS\CM_\CA$).

Each positive sum of points of $\CM_\CA$ is associated with a
representation of the ring $\CA$, which is the direct sum of irreducible
representations. To each element in the ring, we associate a
character in the representation. Consider the vector space associated to
a representation on which the matrix ring acts as a left module of the
ring of functions. Because the associated representations of points are
only well defined up to conjugation, we should impose the same
constraint on the modules; namely, we want isomorphism classes of
modules for the ring $\CA$. This is very reminiscent of algebraic
K-theory, and we will expand on this idea later, making the connection
precise.

Now, although we have found it natural to extend $\CM_\CA$ to
$\CS\CM_\CA$, it is not the case that $\CS\CM_\CA$ automatically
inherits a topology from that of $\CM_\CA$. Rather, we should repeat the
construction of a Zariski topology, by giving a definition of closed
sets. For any function $a\in \CA$ and for every complex number $z$, we
define the following set 
\begin{equation}
\CZ =\{ p \in \CS\CM_\CA | \tr_{R_p}(a) = z\}
\end{equation}
to be closed. These sets form a basis of closed sets for the topology of
$\CS\CM_\CA$. This topology coincides with the natural topology in the
commutative algebra of functions generated by traces of operators, which
are the polynomials in the gauge invariant superfields, and thus gives
us the same topological information that we would want for the moduli
space if we just consider the ring of holomorphic functions on the
moduli space with values in $\BC$. For later use, we define the support
of a character as
\begin{equation}\label{eq:suppchar}
Supp(\tr\ a)=\overline {\left\{ p\in\CS\CM_\CA | \tr_{R_p} a\neq0\right\}}
\end{equation}
where the overline denotes the closure operation.

We would like to be able to say that the space is foliated by sets of
degree $m$ (which will count the $D$-brane charge of the point). Because
the Zariski topology is coarse, we must then additionally declare that
the sets defined by $\deg(x) = m$ are both open and closed. 

Now recall that for two different points in an algebraic variety $V$, 
there is some function on the variety which distinguishes
them. Only a finite number of these functions is needed to determine a
point exactly; the ring of polynomials (with relations) is
finitely generated. This construction is also sufficient to determine a
collection of $n$ unordered points of the variety. By examining
$n$th order polynomials in a function $f$, we can determine the values
that $f$ takes at the $n$ points. If the values are different for all
the points for some function $f$, then one can use $f$ as a coordinate,
and one has a collection of $n$ non-overlapping algebraic subsets of
$V$, with one point chosen from each one. Thus we can construct a
function which vanishes at all but one of the subsets, which we call
$f_1$ and by multiplying $f_1$ by all the basis functions of the
ring associated to $V$, we can identify one of the points. The procedure
can be repeated if no one function is able to tell them all apart, and
then we get the multiplicities of the points.

In the non-commutative case, we say that we can always distinguish two
irreducible representations by some collection of characters (traces) of
the ring $\CA$. Thus there is a given finite number of functions with
which we can distinguish $n$ points.

Recall that we wish to think of the non-commutative symmetric space
as a refined version of the commutative space. Topologically, the two
spaces are the same. It is clear that, at least locally, given the 
characters of enough elements of the ring, we can fully reconstruct 
a representation by holomorphic matrices on the commuting variables.
This endows the symmetric space locally
with a holomorphic vector bundle structure.

\subsection{The Role of the Center}

Now let us apply the above construction to maps between the spaces
$\CS\CM_\CA$ and $\CS\CM_{\CZ\CA}$. Consider in particular the inclusion
map $\CZ\CA\to \CA$, which is the pullback of functions on $\CM_{\CZ\CA}$ to
$\CM_\CA$.

We want to know the image of a point $p$ in $\CM_\CA$. Given the point
$p$, there is an associated $M$-dimensional irreducible representation
$\mu_p$. Consider composing the maps $\CZ\CA\stackrel{\ii}{\to}
\CA\stackrel{\mu_p}{\to} M_M(\BC)$. As the last map is onto, if
$a\in\CZ\CA$, it commutes in the image of the composition of maps and by
Schur's lemma is proportional to the identity. The representation
associated to $p$ splits into $M$ identical copies of a single
representation of $\ZA$, namely, into $M$ copies of a single point. We
write this as
\begin{equation}
p \mapsto M \bar p
\end{equation}
where $\bar p$ is the associated maximal ideal of $\ZA$, which is the
kernel of the inclusion map.

We will call this the {\it natural map}, as it respects degree. 
Notice that we can also define a map
between the symmetric spaces $p \mapsto \bar p$ which forgets the degree. In
this case, the image of a point is a point, so we can restrict the maps
to $\CM_\CA$ and $\CM_{\CZ\CA}$. We will call this the {\it forgetful map}.

The center of the algebra will play an important role in the physics.
We wish to restrict the maps between rings $\CA$ and $\CB$ in the 
following way. Note that we have the following diagrams
\begin{equation}
\begin{matrix} \CA & \rightarrow &\CB\\
\uparrow& &\uparrow\\
\CZ\CA & & \CZ\CB
\end{matrix}\ \ \ \ \ \ \ \ \ \ \ \ 
\begin{matrix} \CS\CM_\CB & \rightarrow &\CS\CM_\CA\\
\downarrow& &\downarrow\\
\CS\CM_{\CZ\CB} & & \CS\CM_{\ZA}
\end{matrix}
\end{equation}
We require that these diagrams be commutative; namely, the ring
homomorphism $\CA\to\CB$ induces a map $\ZA\to \CZ\CB$, and consequently
$\CS\CM_{\CZ\CB}\rightarrow\CS\CM_{\ZA}$. Thus, we want the map to be
such that central elements are central not just in the subalgebra of the
image of $\CA$ but in the algebra $\CB$ itself.

Now we are at a point where we can build a category for the
non-commutative algebraic geometry. The objects will be rings $(\CA)$
with the center identified and the
inclusion map singled out.\footnote{It is appropriate then to use the larger
notation $(\CA) \sim (\CA, \ZA,\ii: \ZA\to \CA)$.} The allowed maps
between rings are such that they produce commuting squares
\begin{equation}
\begin{matrix}\CA & \rightarrow &\CB\\
\uparrow& &\uparrow\\
\ZA & \rightarrow & \CZ\CB
\end{matrix}
\end{equation}
with the upwards arrows the natural inclusion maps.

The non-commutative space is a contravariant functor from this category
of rings to a category of `symmetric spaces' as we have defined previously
(including the degree map and the degree-preserving property). Thus
we have the diagram
\begin{equation}
\begin{matrix} \CS\CM_\CB & \rightarrow &\CS\CM_\CA\\
\downarrow& &\downarrow\\
\CS\CM_{\CZ\CB} & \rightarrow & \CS\CM_{\ZA}
\end{matrix}
\end{equation}
Together with the center preserving property, the forgetful map induces
\begin{equation}
\CM_{\CZ\CB} \to \CM_{\CZ\CA}.
\end{equation}
This is a map of commutative algebraic varieties, to which we can apply
intuition. It is also clear that this map between varieties has all the
data required to specify the map between their symmetric spaces. With
the topology of these spaces, all the arrows are continuous maps, and
composition of maps is a map that respects the properties of the
category. 

To make a full connection with algebraic geometry, we want to be able to
glue rings on open sets. This should be done by a process of
localization. These details will be left for a future publication\cite{BJL}.

It is useful to notice that all the irreducible representations of the
center may not appear when we consider the projection map,
$\CS\CM_\CA\to\CS\CM_{\ZA}$. If most\footnote{That is, an open set of 
$\CM_{\CZ\CA}$ in
the Zariski topology.} do appear, then we will call
the algebra {\it semi-classical}, because to the points in the
variety associated to the center, we can lift to points in
the non-commutative variety. The non-commutative variety covers the
commutative one and this notion will be important from several
perspectives below. In particular, there are applications involving 
$D$-branes in which phenomena on orbifold spaces are more precisely
described by non-commutative geometry.

\subsection{$D$-brane Fractionation}

The technology developed so far contains some interesting aspects of
$D$-brane physics. In particular, we wish to show that a $D$-brane
fractionates as we move to a singular point of a non-commutative
algebraic variety. In fact we define singular points via this process of
fractionation. We will consider in this subsection $D$-branes which
correspond to points in $\CM_\CA$, and the degree of the point is
identified with the $D$-brane charge. The moduli space of supersymmetric
configurations of $D$-branes is identified with $\CS\CM_\CA$.

Let $R$ be an irreducible representation of dimension $M$ in $\CA$.
Consider its image in $\CS\CM_\CA$ as a single point of degree $M$.
Because $\CS\CM_\CA$ is an algebraic variety, it will consist of several
components or branches. The branch of $\CS\CM_\CA$ where $R$ is located
is a closed set of some complex dimension $d$ which is not a closed
subset of any set with larger local dimension.

On this branch, we can define a local function which is the dimension of
the commutant $\CZ R$ of the representation $R$. As we move along the
branch, this function is semi-continuous---it may jump in value on
closed sets.

Clearly, the sets with $\dim(\CZ R)>1$ are closed. In this case, we have
at least two linearly independent matrices which commute with everything
in the image of $\CA$, and thus the representation cannot be
irreducible. For irreducible representations, $\dim(\CZ R)$ must be
unity. Thus, if we start at a point on a branch of the variety that is
irreducible, as we continuously deform along it, we can reach a special
point as a limit point, where the representation becomes reducible.

Parametrize this deformation by $z$; on the symmetric product space we
have the  process
\begin{equation}
\lim_{z\to z_0} x(z) = x_1+\dots + x_n
\end{equation}
if $z_0$ is such a limit point, and  where $n$ is the number of
irreducible representations that $R(z)$ splits into. Then, we say that
the $D$-brane has fractionated, and there may be additional branches that
intersect that point, corresponding to separating the fractional branes.

From the point of view of the center of the algebra, each element is
proportional to the identity throughout the branch of the symmetric
space, and thus there is no splitting seen in $\CS\CM_{\ZA}$. In this
sense, the non-commutative geometry is a finer description of the
$D$-brane moduli space than the associated commutative geometry.

In the cases where there are branches corresponding to separating the
branes at $z_0$, if we think in terms of the forgetful map, we would
have a single point splitting into $n$ points. From the point of view of
the commutative algebraic variety, there is a jump in dimension as we go
from one branch to the next; this is naturally associated with a
singularity. We will see explicit examples of this in Section
\ref{sec:discrete}.

\subsection{Higher dimensional branes}

So far, we have considered $D$-branes that are point-like on the moduli
space. We would also like to identify more general brane configurations,
such as those wrapped on holomorphic subspaces; hence we need to
construct such objects algebraically. It is natural to consider coherent
sheaves\cite{HM}: these are the modules over the ring $\CA$ which
locally have a finite presentation and are well-behaved when considered
from the commutative standpoint. Extended BPS brane solutions usually
correspond to stable sheaves, given some appropriate notion of
stability. Moreover, they are also well-behaved as far as K-theory is
concerned. However, in order to define these structures, it is most
convenient to have a semi-classical ring. This does not mean that there
is no useful way to define these objects for more general rings, but on
some rings where the points are discrete there is no obvious notion of
an extended object. For the rest of this section, we will assume that we
are indeed working in a semi-classical ring.

As the ring is semi-classical, we can try to construct extended objects
by first building them over a holomorphic subspace of the commutative
structure, and then try to lift them up to the non-commutative geometry.
In the commutative case, a $D$-brane corresponds to a coherent sheaf with
support on a commutative subvariety. For any notion of non-commutative
sheaf, it must be the case that it is also a coherent sheaf over the
commutative ring. In the commutative case, the $D$-brane is a module over
the ring $\ZA$, such  that if $\CZ \CI$ is the ideal corresponding to
the support of the sheaf, the module action of $\ZA$ factors through
$\ZA/\CZ\CI$, which is considered to be the coordinate ring of the
closed set associated to the ideal $\CZ \CI$.

On `good' varieties we always have a presentation of a sheaf $\CS$ as
the right-hand term of some exact sequence
\begin{equation}\label{eq:exseqcomm}
\ZA^m \to \ZA^n \to \CS\to 0
\end{equation}
that is, as a module with $n$ generators with relations induced by
the images of $\ZA^m$.

We want to mimic this construction for the non-commutative version
of the $D$-brane. Note that in the non-commutative case, we have a choice
of left-, right- or bi-modules of the algebra $\CA$. However,
physically, we need to consider only bi-modules, as both ends of open
strings end on a $D$-brane. That is, gauge transformations (which act
locally) act both on the left and right, and therefore the algebra
has to be able to accommodate both types of actions on the modules.
Referring to the bi-module as $\CR$, we want them to arise from exact
sequences
 \begin{equation}\label{eq:sheaf}
\CA^m \to \CA^n \to \CR\to 0
\end{equation}
in analogy to eq. (\ref{eq:exseqcomm}). This defines locally\footnote{Recall
that we are not concerned with gluing.} 
the coherent sheaves over $\CA$.

The annihilator of a bi-module $\CR$ is defined as the largest ideal
$\CI$ of $\CA$ such that $\CI \CR = \CR\CI =0$. This is a double-sided
ideal, and thus defines a closed set, in the topology of $\CM_\CA$.
For any point $p$ in $\CM_\CA$
which does not belong to $\CI$, $\CI+\CI_p$ is equal to $\CA$ (because
$\CI_p$ is maximal). We will refer to this ideal $\CI$ as $\Ann(\CR)$.

We can find how a bi-module restricts to a closed subset (described by
an ideal $\CI$) by noticing that if $\CR$ is a bi-module over $\CA$, then
$\CR/(\CI \CR+\CR \CI)$ is a bi-module over $\CA/\CI$. For maximal $\CI_p$ the
restriction is zero if $\Ann(\CR)\notsubset\CI_p$. We
can define the support of a sheaf to be the set of points such that
\begin{equation}\label{eq:support}
\Ann(\CR)\subset \CI_p.
\end{equation}

We study the sheaves locally by restricting to a point. We will look at
two different notions of the rank of a sheaf. Each non-commutative point
is Morita equivalent to a commutative point. This tells us that modules
over the functions restricted to a point $p$ (the ring of $n\times n$
matrices) behave just as vector spaces over $\BC$. Thus the sheaf
restricted to a point is a bi-module over the ring of
$\deg(p)\times\deg(p)$ matrices, and thus $\CR |_p$ is isomorphic to
$\left(\CA |_p\right)^k$, for some $k$. One possible definition of
non-commutative rank of a sheaf at the point $p$  is just $k$. However,
as seen from the commutative standpoint, the dimension of the
representation associated to $\left(\CA |_p\right)^k$ is $k\deg p$, and
this then serves as another definition of rank, which we will refer to
as the commutative rank.

As usual, rank is upper semi-continuous (the points where $rank
(\CR)|_p>m$ form a closed set). The rank can jump in value on some
closed subset, and this is interpreted in terms of an additional $D$-brane
of smaller dimension stuck to the brane, as follows from the anomalous
couplings of $D$-branes\cite{GHM}.

For the  non-commutative points, we have to take into account that a
limit set of a collection of points might be a sum of points. Consider
the trivial bi-module of $\CA$, namely $\CA$. The non-commutative rank
(equal to 1) does not jump under the fractionation process. The
commutative rank on the other hand, does jump at this singularity.
The non-commutative rank then is the natural definition of rank for
non-commutative algebras.

Note, however, that if we look just at the center $\CZ\CA$, the
commutative rank is the natural definition, as it does not jump in a
splitting process; we have
\begin{equation}
\deg(p)  = \sum _i \deg{p_i}.
\end{equation}

With these definitions, a $D$-brane is a coherent sheaf over both
the non-commutative ring and the commutative sub-ring.

\subsection{K-theory interpretation}

We have seen that our approach to non-commutative geometry has led us
to some definitions of $D$-brane states. We now want to add $K$-theory to
the discussion. Because we have a ring and we have bi-modules, we get
automatically a K-theory associated to this structure, namely, the
algebraic K-theory of the ring $\CA$ (see \cite{Ros} for example). From
a mathematical standpoint this is review material, and we will just
glimpse at the dynamics in terms of brane-antibrane systems\cite{Sen,
W3}.

Indeed, let us start with the construction of the symmetric  space. We
had formal sums of points which makes the non-commutative geometry a
semigroup. We can make it into a group by adding minus signs, and a rule
for cancellation. This group is the equivalent of zero-chains of points.

Now, the idea of the group structure is to understand that $p+(-p) =0$.
So if $p$ is a point, we interpret it as a point like $D$-brane, and $-p$
is an anti-$D$-brane. The cancellation law of the group is the statement
that a $D$-brane anti-$D$-brane pair can be created from the vacuum.

Given these minus signs, the degree function now maps to the 
integers and gives us an invariant, which is a  group homomorphism of 
Abelian groups, $\deg : \oplus p\to \BZ$.  This number
is the total  $D$-brane charge of a configuration. It is possible to give
a topology to finite formal sums of points by the same construction we
used based on characters. The extra ingredient to make $p-p=0$ is to add
minus signs for the anti-$D$-brane.

Thus dynamically $p-p=0$ is the statement that any character of $p-p$ is
the same as a character of zero, and thus the configurations are
connected. When we create a $D$-brane anti-D brane pair we can separate
them if there are moduli available, and thus the process is continuous
in this topology. Dynamical information would include the energy
required for this process. (A generic point is such that this energy
is much less than the energy required to move off of the moduli
space of sums of points. A non-generic point is where the mass 
matrix has zero eigenvalues.) 

The idea now is to define the K-theory of points as a homotopy invariant
which respects the additivity of branes. On one hand, we have the
mathematical definition of the $K^p_0$-theory of points as the formal
abelian group of homotopy classes of finite dimensional representations
of the algebra $\CA$, such that if $a,b$ are such representations, then
the K-theory classes associated to $a\oplus b, a, b$ satisfy
\begin{equation}
K(a\oplus b) = K(a)+K(b)
\end{equation}
and if one has a homotopy between the two representations $a\sim b$,
then $K(a)= K(b)$. Indeed, to the point $p$ we associate the bi-module
$\CA|_p$, and this is thought of as the skyscraper sheaf over $p$. With
this extra relation this is part of the K-theory of bi-modules of the
algebra.

The other way to define K-theory is to say $K(a)= K(b)$ if there is $c$
such that $a\oplus c \sim b\oplus c$. Both of these definitions agree.


There are a few possible choices of K-theory depending on the type of
modules one chooses. As we have stated above, in this paper we are
interested in finitely presented bi-modules over the ring $\CA$.
Generically, one defines the K-theory associated to projective
bi-modules of the algebra. The K-theory of projective bi-modules is the
same as the K-theory of finitely presented bi-modules as long as every
bi-module admits a projective resolution (this is true, for example, in
smooth manifolds where every vector bundle is projective over the
coordinate ring of the manifold). Thus, as long as there is a long 
exact sequence
\begin{equation}\label{eq:resolution}
0\to P_1\to\dots P_k \to M\to 0
\end{equation}
with the $P_i$ projective, then the K-theory class of $M$ is defined.
This requires the ring to be regular. Whether or not we will always get
regular rings in string theory in this framework is not clear.
(Singular varieties are not
regular as commutative algebras, yet they do appear in string theory.)

The $K_0$-theory is defined as the set of formal sums of bi-modules
modulo homotopy, and modulo the relations
\begin{equation}
K(b) = K(a) + K(c)
\end{equation}
whenever there is a short exact sequence
\begin{equation}\label{eq:exabc}
0\to a\to b\to c \to 0
\end{equation}
of the bi-modules we have described. We think of this as the statement
that $b-a-c=0$.

If a module $M$ admits a projective resolution as in 
(\ref{eq:resolution})
then it is a simple exercise to show that
\begin{equation}
K(P_1) -K(P_2) +\dots- (-1)^kK(P_k)+(-1)^k K(M) =0
\end{equation}

Physically, we say the dynamics of brane-antibrane configurations
is such that given a short exact sequence (\ref{eq:exabc}),
the process
\begin{equation}
X\to X\pm a\mp b\pm c
\end{equation} 
is allowed, namely $a-b+c$ carries no $D$-brane charge.
In particular, the exact sequence
\begin{equation}
0\to a\to a\to 0\to 0
\end{equation}
will allow any of the two  processes
\begin{equation}
X\to X+a+(-a)\to X
\end{equation}
which correspond to the creation of brane-antibrane pairs.
Of course, the real dynamics of these processes is not available
to us, but the topology of allowed transitions is correctly 
reproduced.

Note that taking tensor products of bi-modules is locally a good
operation (at each point we are taking a tensor product of finite
dimensional spaces, and we get a finite dimensional space). Thus the
K-theory is not just an additive group but we have a multiplication as
well, and this permits us to do intersection theory (that is, we can
count strings when branes intersect). This type of information can often
be enough to calculate topological quantities in string theory.

As a final comment, we have to give some warnings to the reader. The
K-theory we have constructed here is the one associated to the
holomorphic algebra, and thus is a version which is
relevant for the algebraic geometry. This K-theory is an invariant of
a holomorphic space which is much finer than the topological K-theory,
and thus contains a lot more information. The K-theory which is
relevant for $D$-brane charge is the one associated to both the
holomorphic and anti-holomorphic structures, namely, the
K-theory of a $\BC^*$ algebra of which $\CA$ is a subalgebra. This
$\BC^*$ algebra includes the D-term constraints, and by  theorems on
existence of solutions\cite{LT} the geometric space associated to the
$\BC^*$ algebra has just as many non-commutative points as the one
associated to $\CA$. So just as in commutative cases, the
non-commutative holomorphic data parametrize the full variety. The 
K-theory of the two algebras does differ. The holomorphic K-theory is
therefore more appropriate to count BPS states, rather than just account
for the $D$-brane charge.

This is the end of the mathematical preliminaries. We believe that we
have presented a fairly general account of how applications of these
techniques might be pursued. We will see that this approach is not 
just a big machine which describes things we
already knew in a complicated manner. Indeed, once we have examined
the examples in 
the next sections, it will be clear that the formulation brings 
sound intuition and gives a very nice picture of how string
geometry behaves.

%
%
\section{Examples}\label{sec:discrete}

In this section, we consider a variety of 
examples in order to build a picture of the generic behavior of the geometry,
which is not present in the simplest case. The
presentation is given in terms of the language of Section
\ref{sec:algebra}; the reader will find it necessary to read, at least,
the overview in Section \ref{sec:overv}. In the first few examples, we
first calculate the commutative algebra of the center which reproduces
the string geometry associated with the field theory ({\it e.g.}, the
orbifold). We then attempt to build the irreducible representations of
the full non-commutative algebra by exploiting knowledge of the center.
A posteriori, the structures that we find here and the relations
to the physics of $D$-branes in these geometries discussed in later
sections, motivates the formal
constructions of Section \ref{sec:algebra}. 
In more general examples, the calculation of the center is difficult, and we
present only partial results.

\subsection{Orbifolds with discrete torsion: the $q$-deformation}
\label{sec:qdefsec}

Our first example to study will be orbifolds with discrete torsion. In
particular, we consider the orbifold $\BC^3/\BZ_n\times \BZ_n$ with
maximal discrete torsion. To construct the low energy effective field
theory of a point-like brane one can use a quiver construction\cite{DM}
with projective representations of the orbifold group\cite{D}. The use
of projective representations was justified in Ref. \cite{Gomis}. The
algebraic variety associated to the orbifold singularity is given by the
solutions of one complex equation in four variables,
\begin{equation}
x y z = w^n.
\end{equation}
As Douglas showed\cite{D,DF}, the theory has $N=1$ supersymmetry in four
dimensions and consists of a quiver with one node, gauge group $U(M)$,
three adjoint superfields and a superpotential
\begin{equation}
W_q=\tr(\phi_1\phi_2\phi_3) - q \tr(\phi_2\phi_1\phi_3)
\end{equation}
with $q$ a primitive $n$-th root of unity.
This theory can be obtained by a marginal deformation of the $N=4$
supersymmetric field theory as shown in \cite{LS, BL}, and as such, when
studied under the  AdS/CFT correspondence, displays a
duality between two totally different near-horizon geometries, 
describing the same field theory.

The $F$-term constraints are given by
\begin{eqnarray}
\phi_1\phi_2 - q\phi_2\phi_1 &=&0\\
\phi_2\phi_3 - q\phi_3\phi_2 &=& 0\\
\phi_3\phi_1-q\phi_1\phi_3 &=& 0
\end{eqnarray}
We will often write these using the $q$-commutator notation $\qcomm{
\phi_1}{\phi_2}=0$, etc. These equations are exactly the type of
relations seen in the algebras related to quantum planes\cite{Manin},
and have been very well studied. Let us analyze the algebra using the
tools described in Section \ref{sec:algebra}.

Because of the $F$-term constraints, we can always write any monomial in
`standard order'
\begin{equation}
\phi_1^{k_1}\phi_2^{k_2}\phi_3^{k_3}.
\end{equation}
We associate to this monomial the vector $(k_1,k_2,k_3)$.

Note that if an element commutes with $\phi_1, \phi_2,
\phi_3$, then it commutes with any of the monomials, and thus is an
element of the center of the algebra.
Monomials may be multiplied, and up to phases, we have
\begin{equation}
(k_1,k_2,k_3) . (s_1, s_2, s_3) \sim (k_1+s_1,k_2+s_2,k_3+s_3).
\end{equation} 
Because of the phases, generators of the center are
monomials.

It is easy to see that $(k_1,k_2,k_3). \phi_1 = \phi_1.(k_1,k_2,k_3)
q^{k_3-k_2}$, so that $k_3 = k_2 \mod  n$ for $(k_1,k_2,k_3)$ to be
in the center. Similarly one proves
$k_2=k_1\mod n$ and thus the center is given by the condition
\begin{equation}\label{eq:center}
\ZA= \left\{ \sum (k_1,k_2,k_3)\big| k_1 = k_2 = k_3\mod n \right\}
\end{equation}

This is a sub-lattice of the lattice of monomials, and it is generated
by the vectors $(1,1,1), (n,0,0), (0,n,0), (0,0,n)$. Call $w = (1,1,1)$,
$x= (n,0,0)$, $y=(0,n,0)$ and $z=(0,0,n)$. Clearly we have the relation
\begin{equation}\label{eq:orbifold}
(-w)^n + x y z = 0
\end{equation}
so we see the orbifold space is described by the center of the algebra.
The singularities occur along branches where two of $x, y, z$ are zero.

Now that we have the commutative points, let us consider the
non-commutative points of the geometry. We should consider the
irreducible finite dimensional representations of the algebra.
Because $x, y, z$ are central, on an irreducible representation of the
algebra they act by multiples of the identity.

Suppose at least two of $x, y, z$ are non-zero (say $x, y$). In this case
$(1,0,0)$ and $(0,1,0)$ are invertible matrices. By a linear
transformation, we can diagonalize $(1,0,0)$. Consider an eigenvector
$\state{a}$ of $(1,0,0)$ with eigenvalue $a$. We see that
$\state{qa}\equiv(0,1,0)\state{a}$ is an eigenvector of $(1,0,0)$ with
eigenvalue $qa$. Thus we get a collection of states $\state{a},
\state{qa}, \dots ,\state{q^{n-1} a}$ constructed as $\state{a},
(0,1,0)\state{a}, \dots, (0,n-1,0)\state{a}$. This sequence
terminates (and thus the representation is of dimension $n$)
because $(0,n,0)$ is central, and $q^n=1$.
A set of matrices which satisfies these conditions is
\begin{eqnarray}
(1,0,0) = a P\\
(0,1,0) = b Q 
\end{eqnarray}
with $P$ and $Q$ defined by\footnote{We have changed basis compared to
Ref. \cite{D}.}
\begin{equation}\label{eq:PQ}
P = \begin{pmatrix}1&0&0 &\dots& 0 \\ 0& q & 0
& \dots&0 \\ 0& 0& q^2&\dots&0 \\
\vdots&\vdots&\vdots& \ddots&\vdots \\
0&0&0&\dots&
 q^{n-1}\end{pmatrix},\ \ \ \
Q = \begin{pmatrix}0&0& \dots&0&1\\
1&0&\dots&0&0 \\
0&1&\dots&0&0 \\
\vdots&\vdots&\vdots&\ddots&\vdots\\
0&0&\dots&1&0
\end{pmatrix}
\end{equation}
As $w= (1,1,1)$ is central, it is proportional to the identity.
It trivially follows that $(0,0,1) = c  Q^{-1} P^{-1}$.

Notice that our solutions are parameterized by three complex numbers,
namely $a, b, c$. It is easy to see that $x = a^n I$, $y= b^n I$ and $z
= -(-c)^n I$, $w = abc I$, and that one can cover the full orbifold with
these solutions, except for the singularities (where two out of the
three $x,y,z$ are zero). Notice also that the covering is done smoothly,
so any two points can be connected by a path which does not touch the
singularities.

If we label the representation by $R(a,b,c)$, it is easy to see that
$R(a,b,c)$ is equivalent under a similarity transformation to $R(q a,
q^{-1} b, c)$ and $R(q a, b, q^{-1} c)$. Thus the eigenvalues of the
center completely describe the representation. That is, for any
commutative point which is non-singular, we have a unique
non-commutative point of degree $n$ sitting over it.

Let us now analyze the case where two of the three $x,y,z$ are zero.
Then $w=0$ as well, and we are along one of the singular branches of the
orbifold. Let us assume that $x\neq 0$; then $(1,0,0)$ is invertible,
and can be diagonalized. On the other hand $(0,1,0).(0,0,1)=
(0,0,1).(0,1,0) = (0,n,0)= (0,0,n)=0$ in the representation. Given any
vector $v$ in the representation, $v'\equiv (0,n-1,0) v$ is annihilated
by $(0,0,1)$ and $(0,1,0)$, and any other vector obtained by multiplying
with $(1,0,0)$ enjoys this same property. Thus given a representation,
we find a sub-representation where both $(0,0,1)$ and $(0,1,0)$ act by
zero. As $(1,0,0)$ is invertible it can be diagonalized in this
subrepresentation. Clearly the representation is irreducible only if it
is one dimensional, and determined by the eigenvalue of $(1,0,0)$, which
is a free parameter that we call $a$. The value of $x$ is $a^n$, and for
each point in the singular complex line $y=z=0$ we find $n$ irreducible
representations of the algebra, except at the origin. The same result
holds when we go to any of the other complex lines of singularities. We
label these representations by $R(a,0,0)$, etc.

Here $R(a,0,0)$ is not equivalent to $R(qa,0,0)$. They are equivalent as
far as the commutative points are concerned, because both of these
representations have the same characters over the center of the algebra.
But as far as the non-commutative points are concerned, the characters
of the non-central element $(1,0,0)$ differ. That is
\begin{equation}\label{eq:fractional}
\tr_{R(a,0,0)}(1,0,0) = a
\end{equation}
Thus we have two distinct points. It is also clear that any one of these
representations can be continuously connected to any other.

These smaller representations are not regular for the
$\BC^3/\BZ_n\times\BZ_n$ orbifold and may be identified with
the fractional branes. Notice also that
\begin{equation}
\tr_{R(a,b,c)}(1,0,0) = a \tr P = 0
\end{equation}
so that this character is different from zero only at the classical
singularity. This is the primary reason for adopting the convention for
the support of a character in eq. (\ref{eq:suppchar}).

To summarize, for each point in the classical moduli space we have at least one
point in the non-commutative space which sits over it. This is an
example of a {\it semi-classical} geometry (see Section
\ref{sec:algebra}). The commutative singular lines are covered by an
$n$-fold non-commutative complex plane branched at the
origin.

Now consider what happens when we bring a point from the regular part
of the orbifold towards the singularity. The
representation behaves in this limit as
\begin{equation}\label{eq:decompose}
\lim_{b, c\to 0} R(a,b,c) = 
R(a,0,0)\oplus R(qa, 0,0)\oplus\dots R(q^{n-1}a, 0,0)
\end{equation}
In our description of moduli space, this corresponds to the branes
becoming fractional at the orbifold fixed lines, as we have discussed
previously. Indeed, once we reach this point we can separate the
fractional branes, and the non-commutative symmetric product is the
right tool for describing the moduli space in full.

We can also see the quiver of the singularity type by consideration of
this same limit. Indeed, we assign a node to each irreducible
representation in the right-hand side of eq. (\ref{eq:decompose}).
We draw an arrow between any two nodes appropriate to the non-zero
entries in eqs. (\ref{eq:PQ}) and we obtain Figure
\ref{fig:quiver} which is indeed the quiver diagram of the orbifold in
the neighborhood of a point in the singular complex line.
\FIGURE[2]{
\littlefig{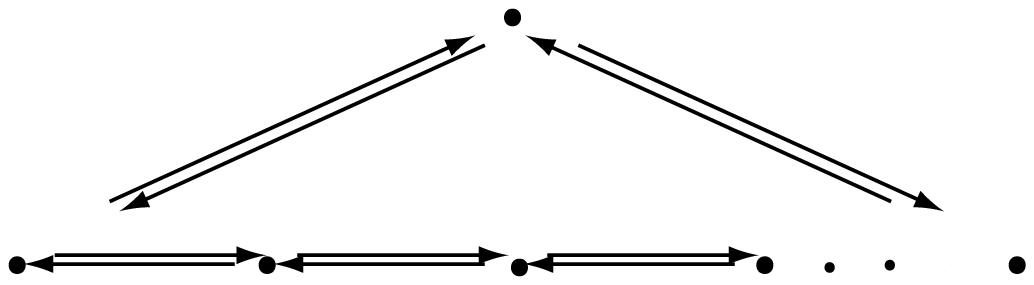}{3}
\caption{Quiver diagram for the $A_{n-1}$ singularity.}
\label{fig:quiver}}
Thus the singularities can be said to be locally quiver. 
From the point of view of the center of the algebra, the nodes of
the quiver are at the same point, but they are distinct in the
non-commutative algebra. The behavior of the field theory near the
singularities is precisely what we would get from the orbifold analysis.

Recall that the commutative singular lines are covered by $n$
non-commutative branches. The monodromies of the quiver diagram are 
encoded in this structure, and thus their calculation is geometrically
obvious. This compares quite favorably to the rather cumbersome procedure
employed in \cite{BL}.
Indeed, we can change $a\to\omega a$ for $\omega=e^{2\pi i/n}$.
This results in a permutation of the factors appearing on the right-hand
side of eq. (\ref{eq:decompose}). This permutation is the monodromy.

\subsection{Adding one mass term}

Next, we consider a relevant deformation of the last theory, obtained
by the addition of a single mass term. This theory is
a $q$-deformed version of the theory which flows in the infrared
to an $N=1$ conformal field theory\cite{LS,AILS}.
The superpotential is
\begin{equation}
W=\tr\left(\phi_1\phi_2\phi_3 - q\phi_2\phi_1\phi_3
 + \frac {m}2 \phi_3^2\right)
\end{equation}
Again, we assume that $q$ is an $n$-th root of unity.
The $F$-term constraints are given by
\begin{eqnarray}
\qcomm{\phi_1}{\phi_2} &=& -m\phi_3\\
\qcomm{\phi_2}{\phi_3} &=& 0\\
\qcomm{\phi_3}{\phi_1} &=& 0
\end{eqnarray}
As in the previous case, we look for the center of the algebra 
to obtain the commutative manifold.
It is easy to see that $z = \phi_3^n$ is still in the center.
We can also show that
\begin{eqnarray}
[\phi_1^n,\phi_2]& =& \phi_1^n\phi_2 - q\phi_1^{n-1}\phi_2\phi_1
+q\phi_1^{n-1}\phi_2\phi_1- q^2\phi_1^{n-2}\phi_2\phi_1^2+\dots\nonumber\\
&=& \phi_1^{n-1} (-m\phi_3)+ q\phi_1^{n-2}(-m\phi_3 )\phi_1+\dots\\
&=& -m \phi_1^{n-1}\phi_3\sum_{r=0}^{n-1}q^{2r}\nonumber
\end{eqnarray}
which vanishes, apart from at the special values $q=\pm 1$. Similarly
one proves that $\phi_2^n$ is central, away from $q=\pm 1$. For now, we
will assume that $q^2\neq 1$, and return to these cases later. Thus we
have at least three central variables $x = \phi_1^n$, $y=\phi_2^n$,
$z=\phi_3^n$.

The variable $w$ is modified by the presence of the mass term. 
Consider the commutator
\begin{eqnarray}
[\phi_1\phi_2\phi_3,\phi_1] &=& \phi_1\phi_2\phi_3\phi_1 - 
q\phi_1\phi_2\phi_1\phi_3+q\phi_1\phi_2\phi_1\phi_3 - 
\phi_1\phi_1\phi_2\phi_3\nonumber\\
&=& m \phi_1\phi_3^2\nonumber
\end{eqnarray}
This result may be rewritten as a commutator for $q\neq\pm 1$, and
thus we see that
\begin{equation}
w = \phi_1\phi_2\phi_3 + \frac{m}{1-q^2} \phi_3^2
\end{equation}
is central.

The four variables $x, y, z, w$ are related by
\begin{equation}
x y z = -(-w)^n - \left(\frac{m}{1-q^2}\right)^n z^2
\end{equation}
This is a deformation of the complex structure of (\ref{eq:orbifold}). It
is easy to see that we now have singularities at $w= xz = yz = x y + 2t
z=0$ with $t = (m/(1-q^2))^n$. Thus, the singularities are at $xy = 0,
w=0, z=0$, and so we have two lines of singularities $x=0$ and $y=0$.
The mass term has resolved one of the three complex lines of
singularities (for $q^2\neq 1$).

It is easy to check that a general solution is of the form
\begin{eqnarray}
\phi_1& = & a P \\
\phi_2 &=& -b P^{-1} Q -c P^{-1} Q^{-1}\\
\phi_3 &=& d Q^{-1}
\end{eqnarray}
with $P,Q$ defined as in (\ref{eq:PQ}),  and where 
$a, b, c, d$ are numbers satisfying 
\begin{equation}
ac(1-q^2) = m d 
\end{equation}
One then gets $x = a^n I$, $y= -(b^n+c^n)I$, $z = d^nI$ and $w = -abdI$.
Note that this representation has been chosen such that $\phi_1$ is
diagonal at the singularity $y=0, z=0$.

Because we have a three complex parameter solution of the equations, we
at least cover an open patch of the commutative variety, and we are
again in a semi-classical ring. Indeed, we cover everything by finite 
matrices except $x=0$, as then $c$ is infinite.

A patch which does cover $x=0$ is given by
\begin{eqnarray}
\phi_1& = & -a PQ^{-1} - c P Q\\
\phi_2 &=& b P^{-1} \\
\phi_3 &=& d Q^{-1}
\end{eqnarray}
with $ab(1-q^2)=mq d$. This will be a good description for $y\neq 0$.
The two patches cover the two lines of singularities.
There is still the closed set $x=y=0$ which is not covered by either
patch. We can find solutions for this set by taking $\phi_3$ diagonal
and making an ansatz for $\phi_1$ which is upper triangular with 
entries just off-diagonal and $\phi_2$ a similar lower triangular matrix.
The dimension of this representation is also $n$ and depends on one
complex parameter, namely the eigenvalues of $\phi_3$. 

On approaching the singularity $y=z=0$ from the bulk, we again get a
split set of irreducible representations as follows:
\begin{equation}
\lim_{b,d\to 0} R(a,b,d) = R(a,0,0)\oplus  R(qa,0,0)\oplus \dots
\end{equation}

\subsubsection{Comments on the Infrared CFT}

This case is also very interesting from the field theory perspective
because by adding one mass term to a theory with three adjoints, we
obtain a nontrivial conformal field theory in the infrared.

On the moduli space, we have the following $U(1)$ symmetries
\begin{eqnarray}
a, b, c &\to& \lambda \gamma a, \lambda \gamma^{-1} b, \lambda\gamma^{-1} c\\
d&\to&\lambda^2 d
\end{eqnarray}
where we refer to the parameterization for $x\neq0$. The transformation
given by $\gamma$ is an ordinary $U(1)$, while $\lambda$ is the $U(1)_R$
symmetry, that of the superpotential in the infrared. The $R$ charges
can be chosen in such a way that the superpotential is invariant at the
infrared fixed point. Indeed, one can integrate out $\phi_3$ and one
finds a theory in the infrared with a quartic superpotential
\begin{equation}
\frac{1}{2 m} \tr (\phi_1\phi_2-q\phi_2\phi_1)^2
\end{equation}
This superpotential is a marginal deformation of the infrared theory.
Note that we also have, in the infrared, a $\BZ_2$ symmetry
$\phi_1\leftrightarrow\phi_2$ which ensures that the anomalous
dimensions of $\phi_1$ and $\phi_2$ are equal. (In the ultraviolet, this
$\BZ_2$ symmetry is absent, as we would also have to simultaneously
exchange $q\to q^{-1}$ and rescale $m$.) This symmetry exchanges the two
singular complex lines of the commutative moduli space.

\subsubsection{Special Cases: $q=\pm1$}

Let us return to discuss the moduli space for the cases $q=\pm1$ from
the algebraic point of view.

For $q=1$, which is the $N=4$ theory with one mass term, the moduli
space is the set of solutions to
\begin{equation}
[\phi_1,\phi_2] = -m\phi_3
\end{equation}
with all other commutators vanishing. For an irreducible representation,
$\phi_3$ is central and thus a constant. Because the commutator of
$\phi_1,\phi_2$ is a constant, we get the Heisenberg algebra, and the
only finite dimensional representations are those with $\phi_3=0$. Thus the
moduli space is a commutative space consisting of the symmetric product
of the complex plane, $\BC^2$. Notice that this space is of complex
dimension two and not complex dimension three as in the generic case
studied above. Indeed, in this case the center of the algebra is
generated by $\phi_3$. Because $\phi_3=0$ on the moduli space, we can
actually relax the condition for an element being central: we can take
$\phi_1,\phi_2$ as central elements, which makes the moduli space commutative.

Indeed, this is a case where the algebra is not semi-classical. The
variety associated to the center is the algebra of $\BC$. The
non-commutative space is $\BC^2$, which projects to the origin of $\BC$.
The two have almost nothing in common.

As far as the commutative variety is concerned, the moduli space is a
point. Notice that in this case when we integrate out the field $\phi_3$
we get the correct dimension of the moduli space by counting fields.
This does not happen for generic $q$.

For $q=-1$, we can find the two dimensional solution
\begin{equation}
\phi_1= a\sigma_1,\phi_2 = b\sigma_2, \phi_3 = 0
\end{equation} 
plus two one-dimensional branches where either $\phi_1$ or $\phi_2$ 
is zero.

Here, the center is generated by $\phi_3^2$. Indeed, it can be shown
that these solutions exhaust the list of irreducible representations of
the $q=-1$ algebra. This result follows from the fact that $[zx,y]\sim
z^2$, so a finite dimensional representation must have
$z^2=0$.\footnote{Nilpotent possibilities, such as $\phi_3 \sim
\begin{pmatrix}\ 0 & a\\ 0 &0\end{pmatrix}$ are ruled out by $D$-terms.}

The lesson to be learned from these special examples is that the
commutative and non-commutative spaces may contain little or no
information about  each other when the center of the associated algebra
is small. In this case, the full algebra is an infinite dimensional
vector space over the center, and by considering only finite
dimensional representations, we miss a lot of information.

\subsection{One mass term and a linear term}
\label{sec:masslin}

We can easily modify the previously studied cases by adding a linear
term to the superpotential
\begin{equation}
W=\tr\left( \phi_1\phi_2\phi_3 - q\phi_2\phi_1\phi_3 +
\frac m2\phi_3^2+\zeta_3\phi_3\right)
\end{equation}
Note that by a field redefinition of $\phi_3$,
this is equivalent to adding a mass term
\hbox{$\frac{\zeta_3}{m}(q-1)\tr\phi_1\phi_2$}. We will see that the usual
intuition for mass terms fails in this case, namely, that the moduli
space is not destroyed by the quadratic terms. On the other hand, if we
had added $\tr\phi_1\phi_2$ for $q=1$, we would indeed expect the space
of vacua to be reduced to a set of points.

It is straightforward to show that  $x= \phi_1^n, y=\phi_2^n$, and $z=\phi_3^n$ 
are central, and that
\begin{equation}
w = \phi_1\phi_2\phi_3 +\frac {m}{1-q^2}\phi_3^2  +\frac{\zeta_3}{1-q}\phi_3
\end{equation}
is also central, provided that $q\neq \pm 1$.

The relation between the central elements is
\begin{equation}
xyz = -(-w)^n - \left(\frac {m}{1-q^2}\right)^n z^2+ 
\left(\frac{\zeta_3}{q-1}\right)^n z
\end{equation}
and a generic solution of the equations is provided by
\begin{eqnarray}
\phi_1 &= & a P\\
\phi_2 &=& -b P^{-1} Q + c P^{-1}- d P^{-1} Q^{-1}\\
\phi_3 &=& e Q^{-1}
\end{eqnarray}
with $a c (1-q) = -\zeta_3$, $a d(1-q^2) = m e$, and $x= a^n$, 
$y = -b^n+c^n-d^n$,
$z = e^n$, $w = -abe$.
The singularities now occur at $z =  w = 0$ and 
\begin{equation}\label{eq:xycon}
x y  = \left(\frac{\zeta_3}{q-1}\right)^n
\end{equation}
The two complex lines of singularities that met at the origin when
$\zeta_3=0$ are now replaced by a single $\BC^*$, a cylinder. In the
parameterization above, this corresponds to $b, d, e=0$.
We see that the
non-commutative $\BC^*$  is an $n$-fold cover of the cylinder without
branch points, and again the monodromies of the cover are manifest,
since we chose $\phi_1$ diagonal. This is again a semi-classical ring.

In addition, there are finite dimensional representations which may be
thought of as deformations of $SU(2)$ representations. These occur for
$x=y=0$ and cover regions not captured by the parameterization above.
Some solutions give rise to isolated fractional branes at $x=y=0$.
A similar effect occurs in Section \ref{sec:thrmass} and we will return to a 
full discussion there.

The values $q=\pm1$ are special, as in previous cases, in the sense
that singularities occur, and the non-commutative algebra is not 
semi-classical.

\subsection{Three linear terms}

Consider the superpotential
\begin{equation}
W = \tr (\phi_1 \phi_2 \phi_3) - q\, \tr (\phi_2 \phi_1 \phi_3) + 
\sum_i (q-1) \zeta_i\, \tr \phi_i.
\label{eqn.3F}
\end{equation}
This case was studied in Ref. \cite{D} using gauge invariant
variables. Our conclusions will be consistent with that analysis.

For convenience, we have rescaled the $\zeta$ parameters by a factor of
$(q-1)$. The $F$-terms give
\begin{equation}
\qcomm{\phi_1}{\phi_2} = (1-q)\zeta_3,\ \ \ \
\qcomm{\phi_2}{\phi_3} = (1-q)\zeta_1,\ \ \ \
\qcomm{\phi_3}{\phi_1} = (1-q)\zeta_2
\end{equation}

A possible parameterization is
\begin{equation}
\phi_1 = aP - \frac{\zeta_3}{b} Q^{-1}P,\ \ \ \ 
\phi_2 = -b P^{-1}Q + \frac{\zeta_1}{c}Q,\ \ \ \
\phi_3 = c Q^{-1} + \frac{\zeta_2}{a} P^{-1}
\end{equation}
Note that $x_1=\phi_1^n$, $x_2=\phi_2^n$ and $x_3=\phi_3^n$ are central,
while the fourth central variable takes the form
\begin{equation}
w  =  \phi_1 \phi_2 \phi_3 - \zeta_1 \phi_1 - q \zeta_2 \phi_2 - \zeta_3 \phi_3
\end{equation}
In the given basis, we find $x_1=a^n-(\zeta_3/b)^n$,
$x_2=-b^n+(\zeta_1/c)^n$, $x_3=c^n+(\zeta_2/a)^n$ and $-w= abc + q
\frac{\zeta_1 \zeta_2 \zeta_3}{abc}$.

These four variables are related on the moduli space by
\begin{equation}
x_1x_2x_3 - \sum_i \zeta_i^n x_i + 
2 \beta^n T_n \left( -\frac{w}{2\beta} \right) = 0,
\label{eqn.3F.mod-sp}
\end{equation}
where $\beta \equiv (q \zeta_1 \zeta_2 \zeta_3)^{1/2}$ and $T_n(x) = \cos
(n\cos^{-1} x)$ is the $n$-th Chebyshev polynomial of the first kind.

\subsection{Three mass terms}\label{sec:thrmass}

Next, we consider a rank 3 mass term of the form
\begin{equation}
W = \tr (\phi_1 \phi_2 \phi_3) - q\, \tr (\phi_2 \phi_1 \phi_3) + 
\frac{1}{2}m \sum_i \tr \phi_i^2.
\label{eqn.3m}
\end{equation}
This superpotential has a $\BZ_2\times \BZ_2$ symmetry that changes two
of the $\phi_i\to -\phi_i$, and a $\BZ_3$ cyclic symmetry that permutes
the $\phi_i$. This is the remnant of the $SU(4)_R$ symmetry
group of the $N=4$ SYM theory. The group generators do not commute with
each other, and this symmetry is enhanced to $SU(2)$ when $q=1$. Thus the
symmetry is a subgroup of $SU(2)$ which contains a $\BZ_2\times \BZ_2$
and a $\BZ_3$ subgroup. These are the symmetries of the tetrahedron,
$\hat E_6$, and
since they arise from the $SU(4)$ R-symmetry they are chiral.

This superpotential yields the $F$-flatness conditions (cyclic on $j$,
mod 3)
\begin{equation}\label{eq:qcomm3F}
\qcomm{\phi_j}{\phi_{j+1}} = \phi_{j+2},
\end{equation}
where we have rescaled the fields in order to eliminate a factor of $m$.

We wish to find representations of this algebra; we will not immediately
assume that $q$ is a root of unity. There is a certain class of solutions
which may be thought of as deformations of representations of $SL(2,\BC)$.

Note that (for $q\neq1$)
there is a one-dimensional representation
\begin{equation}
\phi_j=\frac{1}{1-q}
\end{equation}
Higher dimensional representations may always be constructed as
$\phi_i=\frac{1}{1-q}I$, but this is clearly reducible. An irreducible
2-dimensional representation (for $q\neq -1$) is given by
\begin{equation}
\phi_j = \frac{-i}{q+1}\sigma_j, 
\end{equation}
where the $\sigma_j$ are the Pauli matrices.  We can construct higher
dimensional irreducible representations by making the following ansatz:
we suppose that one of the fields, $\phi_3$, is diagonal and traceless, 
and that the
other two fields only have non-zero elements just off the diagonals.
(For $q=1$, these reduce to standard $M$-dimensional $SL(2,\BC)$
generators). We have not been able to construct a proof that all such
irreps may be obtained this way. These are the representations which
respect the discrete chiral symmetry of the system, and are all obtained
from the deformation of the representations of $SL(2,\BC)$. The eigenvalues
will thus be paired $\pm \alpha_k$ and will be the same for all three
matrices because the symmetries are respected.

The explicit forms for the representation matrices fall into two
classes, with dimensions $M=2p$ and $M=2p+1$, the analogues of
half-integer and integer spins.

\newcommand{\dimq}[2]{\sigma_{#2}[#1]}

For $M = 2p$, one finds
\begin{eqnarray}
(\phi_1)_{k\ell} & = & \delta_{k+1,\ell} a_k + \delta_{k-1,\ell} 
\frac{b_{\ell}}{a_\ell}, \\
(\phi_2)_{k\ell} & = & iq^{k-p} \delta_{k+1,\ell}\,  a_k - 
iq^{p-k+1} \delta_{k-1,\ell}\,  \frac{b_{\ell}}{a_\ell}, \\
(\phi_3)_{k\ell} & = & i\alpha_k \delta_{k\ell}
\end{eqnarray}
and we have $b_{p+j}=b_{p-j}$ for $j=1,2,\ldots, p-1$, and
$\alpha_{p+n}=-\alpha_{p-n+1}$ for $n=1,2,\ldots, p$.

The $a_k$'s may all be set to, say, unity, by $SL(M)$ transformations.
The $b_j$'s are determined recursively by
the formula\footnote{We've defined $\dimq{q}{x}=1+q+q^2+\ldots+q^x$.}
\begin{equation}\label{eq:recev}
b_j = \frac{-\frac{q}{1+q}\dimq{q}{2(p-j)} + 
b_{j-1} \left( 1 + q^{2(p-j)+3} \right)}
{q^2 \left(1 + q^{2(p-j)-1} \right)};\;\;\; b_0 = 0,
\end{equation}
for $j=1,2,\ldots, p-1$.
The recursion relation is solved by
\begin{equation}\label{eq:rec1}
b_{k,p} = \frac{q(q^{4p}-q^{2k})(q^{2k}-1)}{(q^2-1)^2(q^{2p}+
q^{2k-1})(q^{2p}+q^{2k+1})}
\end{equation}
and notice that all singularities (poles and zeroes) happen for 
$q$ a root of unity.

All three matrices have the eigenvalues
\begin{equation}
\pm \alpha_n = \pm \frac{1}{q^{p-n}(1+q)}\dimq{q}{2(p-n)}.
\end{equation}
for $n=1,2,\ldots,p$.

When $M=2p+1$, we have instead
\begin{eqnarray}
(\phi_1)_{k\ell} & = & \delta_{k+1,\ell}\, a_k + 
\delta_{k-1,\ell} \frac{b_\ell}{a_\ell}, \\
(\phi_2)_{k\ell} & = & i q^{k-p-1/2} \delta_{k+1,\ell}\, a_k - 
iq^{p-\ell+1/2} \delta_{k-1,\ell}\,  \frac{b_\ell}{a_\ell}, \\
(\phi_3)_{k\ell} & = &i\alpha_k\delta_{k\ell} 
\end{eqnarray} 
where $b_{p+n}=b_{p-n+1}$ and
\begin{equation}
b_n = \frac{-q \dimq{q^2}{p-n} + b_{n-1} \left( 1 + q^{2(p-n+2)} \right)}
{q^2 \left(1 + q^{2(p-n)} \right)};\;\;\; b_0 = 0,
\end{equation}
for $n=1,2,\ldots,p$. The recursion relation is solved by
\begin{equation}\label{eq:rec2}
b_{k,p} = \frac{q(q^{2k}-1)(q^{4p}-
q^{2k-2})}{(q^2-1)^2(q^{2p}+q^{2k-2})(q^{2p}+q^{2k})}
\end{equation}
and again we see that all singularities happen for roots of unity.
We also have $\alpha_{p+r+1}=-\alpha_{p-r+1}$ for 
$r=0,1,\ldots,p$ and the eigenvalues of each matrix are in this case 
\begin{equation}
0,\, \pm \alpha_n = \pm \frac{\dimq{q^2}{p-n}}{q^{(M-2n)/2}}
\end{equation}
for $n=1,2,\ldots, p$.

Note that the solutions that we have written here are not $D$-flat. However,
by standard theorems, there exists such a solution, which is an $SL(M)$
transformation of the stated solutions. Still, we must be careful in drawing
conclusions based on these solutions. In particular, there are apparent
singularities at special values of $q$. We will analyze this point 
further in Section \ref{sec:intsing}.

\subsubsection {Finding more solutions}

So far, we have found representations of the algebra which in the limit
$q\to 1$ reduce to finite dimensional representations of the $SL(2,\BC)$
algebra. We also noted an additional one-dimensional representation
which becomes singular in this limit, and therefore corresponds to a
vacuum of the theory, which goes to infinity in the limit.
This additional solution is characterized by the property $\tr\phi_1\neq
0$, whereas for all the other solutions $\tr\phi_1=0$. 

We should ask if there are more irreducible representations of this
algebra, that we have not found above.
The answer must be yes, because for $q\to -1$ many of the solutions which
correspond to irreducible representations of $SL(2,\BC)$ go away to infinity
(the eigenvalues of the matrices are rational functions of $q$ with
finite numerator and $q+1$ in the denominator. Thus they are infinitely
far away in field space, and do not describe vacua of the theory.)

We can construct additional irreps that do not disappear in the 
$q\to-1$ limit as follows.
The discrete subgroup of $SU(2)$ has a three dimensional representation
in terms of Pauli matrices, which suggests the following Ansatz for the
representations.

The following satisfy the algebra (\ref{eq:qcomm3F})
\begin{eqnarray}
\phi_1 &=& \phi_1'\otimes (-i\sigma_1)\\
\phi_2 &=& \phi_2'\otimes (-i\sigma_2)\\
\phi_3 &=& \phi_3'\otimes (-i\sigma_3)
\end{eqnarray}
if we have
\begin{equation}
\left[\phi_j',\phi_{j+1}'\right]_{-q} =\phi_{j+2}'.
\end{equation}
Thus, if we know solutions for a given $q$, we generate solutions for
$-q$ in this way. These representations are reducible. We will refer
to the the irreducible representations obtained in this way as twisted.
There are
two cases to consider, `half integer' spin and `integer spin'
representations.

The integer spin representations have each eigenvalue repeated twice,
including zero and  are  split into two irreducible representations with
eigenvalues for $\phi_3$ in the succession
\begin{equation}
\pm i \alpha_1\to \mp i \alpha_2\to 
\pm i\alpha_3\to\dots\to 0 \to \dots\to \mp i \alpha_2\to \pm i \alpha_1
\end{equation}
These satisfy $\tr(\phi_3)\neq 0$, and $\tr\phi_{1,2}=0$, as these are
off-diagonal. The broken $\BZ_2$ exchanges these two representations. By
acting with the $\BZ_3$ symmetry we get a total of six new representations
for each even-spin irreducible representation of $SU(2)$.

The `half-integer' cases satisfy $\tr(\phi_{1,2,3})\neq 0$. One can
clearly see a splitting into two irreducible representations, but
because there is no eigenvalue $0$, this splitting into two is reducible
and in total  we get four new representations of the algebra. One of these
is a $\BZ_3$ singlet, and the other three form a triplet.

\subsubsection {Interpreting the singularities}
\label{sec:intsing}

In this section, we will study properties of representations. In general
there are two classes of representations, irreducible and reducible. In
the reducible case, there is no mass gap (classically) as some part of
the gauge group is unbroken (apart from the decoupled $U(1)$). The case
of irreducible representations are potentially more interesting as they
confine magnetic degrees of freedom. We will exploit $S$-duality to find
dual configurations that are electrically confining. Note that as we have
not been able to prove that all irreducible representations are accounted
for, we cannot be sure that we see all of the vacua. For the sake of the
present argument, we will assume that the classification is complete and
try to extract conclusions about the non-perturbative behavior of the theory.

The representations we have found are all matrices which are rational
functions of $q$. From the solutions 
(\ref{eq:rec1}),(\ref{eq:rec2}), we see
that there are poles at roots of unity, $q^n = 1$. 

In the case where we have zeroes and not poles, one observes that as we
take the limit to an appropriate root of unity, the matrix decomposes
in block-diagonal form. Thus the representation becomes reducible in the
limit, and we get various copies of the same type representations (of
lower dimension). These singularities are interpreted in the field
theory as having enhanced gauge symmetry, because the commutant of the
representation is larger. 
If one pictures the vacua of fixed rank as a covering of the $q$-plane, 
we have branch points at some roots of unity.

There are other singularities at roots of unity in the denominators of
the fields $\phi_{2,3}$. As these are not singularities in the
eigenvalues of the matrices, it is not clear that these are singular
solutions. This may correspond to an unfortunate choice of basis for the
representation.

Considering that the roots of unity are special, in the sense that they
are related to orbifolds with discrete torsion which have a very nice
semi-classical geometry associated to them, and also considering that in
the limit $q\to \pm 1$ an infinite family of solutions to the vacua
disappear (in this case there are singularities in the eigenvalues of
the matrices), it is plausible that these are actually bona-fide
singularities and the vacua go to infinity. As we will see, at these 
values of $q$, there are moduli spaces of vacua and this is how we
interpret the singularities.

Let us begin with a discussion of $q=\pm1$.
First, we know that at $q=1$ all of the states which break the chiral
symmetry disappear. Thus we get a jump in the Witten index at this
special value. It is also the case that here for some representations
one sees no signal of the eigenvalues of the matrices being badly
behaved, but it is true that we get poles in the off-diagonal elements.

Let us now discuss $q=-1$, paying particular attention to discrete
chiral symmetry breaking. For $U(M)$, $M$ even, the $q$-deformed $SU(2)$
representations move off to infinity at $q=-1$, and thus all the irreducible
representations come from the `half integer' twisted case.
Thus the Higgs vacua break the $\BZ_2\times \BZ_2$ subgroup completely,
and the vacuum has  an unbroken $\BZ_3$ subgroup. Each of the four
vacua have the $\BZ_3$ embedded differently.

For $U(M)$, $M$ odd, there are irreducible representations
of either integer or half-integer twisted type.
Thus, some of the Higgs vacua break the group to 
an unbroken $\BZ_3$ as in the previous case, and some leave an unbroken
$\BZ_2$ if they are constructed from the `integer spin' type
representations. In addition, the $q$-deformed representations survive
for $q\to-1$ but are reducible (the matrix elements $b_{k,p}\to0$).

Notice that in the previous arguments we have used only the perturbative
symmetries of the theory. We believe that the $SL(2,\BZ)$ S-duality of
$N=4$ SYM is realized and perhaps enlarged in the present case in some
way. We will not address that interesting question here; instead, we
confine ourselves to a few remarks based on $SL(2,\BZ)$ alone.

Because of the $SL(2,\BZ)$ symmetry, at the $N=4$ point we can make a
map of gauge invariant operators between the different dual theories.
Thus we can follow the deformations of the theory for any S-dual
configuration of the $N=4$ theory we start with.

Because of the symmetries preserved by the superpotential, changing from one dual picture
to another keeps the general form of the Lagrangian invariant. Thus we
have a map between couplings $(g,q)\to (g',q')$, and $m\to m'(g,q)$.
Because at the roots of unity the theory is special (many vacua
collide), the roots of unity must be preserved by the S-duality action
on the space of field theories, thus the most general holomorphic
transformation that keeps $q=1$ fixed and the structure of the
singularities is of the form $q\to q^{\pm 1}$.

Given a vacuum that disappears at a root of unity, let's say a Higgs
vacuum, any of the vacua related to it by S-duality also disappear. For
$q=-1$ and $M$ even, the trivial vacuum is $S$-dual to the $q$-deformed
Higgs vacuum which moves off to infinity as $q\to-1$, and thus the trivial
vacuum is also removed. For $M$ odd, again the trivial vacuum
is $S$-dual to the $q$-deformed Higgs vacuum. The latter is reducible, and thus
does not appear to have a mass gap; we conclude that the trivial vacuum is
not confining.  There are still the twisted representations, and thus at $q=-1$,
confinement implies (discrete) chiral symmetry breaking.

If $q$ is a more general $n$th root of unity, even though we get poles in the
$b_{k,p}$, we have not been able to find any gauge invariant chiral
quantity which becomes singular. This suggests that the poles are
obtained from a coordinate singularity. In any case, there seems to be an upper
bound on the number and dimension of irreducible representations, as each of
these general representations seems to decompose into irreducibles of
smaller rank. The bound is given in terms of $n$.

Thus as $q$ goes to a root of unity, we can obtain enhanced gauge
symmetry. If we do an S-duality transformation and use some more
general combination of electric and magnetic
condensates, there will still be an upper bound on the
dimension of irreducibles and thus no mass gap. 

The upper bound on the irreducibles also suggests that one can construct
a large center for the algebra. Indeed, one can take the direct sum of
all the irreducible representations of the algebra we have constructed.
If there are no more irreducible representations, this is a finite
dimensional reducible representation, and the subalgebra of the
$\phi_i$ which is the inverse image of the center of the representation
is a large center for the full algebra.

Experience with the example in Section \ref{sec:masslin} suggests that
in this case one might actually get a moduli space of vacua. As we have
argued that we get a finite number of discrete vacua, let us now show
that there is a moduli space for roots of unity $q^n=1$ with $n>2$.

We would want the moduli space to be built out of the $P,Q$ matrices in
some simple fashion. Let us choose $\phi_1$ to be diagonal. Without
the mass deformation, the $\phi_i$ contain $P,Q,P^{-1} Q^{-1}$. Indeed,
one can see that only with the powers $P^{\pm 1}$, $Q^{\pm 1}$ can one
get a single factor of $q$ in the commutation relations, and there is a
potential to get a cancellation of terms. Thus we take
\begin{equation}
\phi_1 =  a_1 P + a_2 P^{-1}
\end{equation}
Because of the symmetry between $P,Q$, we also take
\begin{equation}
\phi_2 = a_3 Q + a_4 Q^{-1} 
\end{equation}
and the $q$ commutation relations are as follows
\begin{equation}
\phi_1\phi_2- q\phi_2\phi_1 \sim PQ^{-1} + QP^{-1}
\end{equation}
so we take
\begin{equation}
\phi_3 = a_5 P Q^{-1} + a_6 QP^{-1} 
\end{equation}
The parameters are related by
\begin{eqnarray}\label{eq:ai}
a_1 a_4 (1-q^2) &=& m a_5 \\
q^{-1} a_5 a_2 (1-q^2) &=& m a_4
\end{eqnarray}
and thus it follows that
\begin{equation}\label{eq:aimq}
 a_5 a_6 = a_3 a_4 = a_1 a_2 =  \frac{q m^2}{(1-q^2)^2}
\end{equation}
so apart from factors depending on $m,q$, $a_{2i+1} a_{2i} \sim 1$. This
cuts the number of variables from six down to three, and (\ref{eq:ai})
provides one more constraint. Thus we are left with a two parameter
solution of the $F$-term constraints. More surprisingly, these also
solve the $D$-term constraints. These representations are inequivalent 
as one can show
that the gauge invariant vacuum expectation value $\tr(\phi_1^n)$ is not
independent of the $a_i$. For $q= \pm 1$, eq. (\ref{eq:aimq}) shows that the
$a_i$ are singular, and  $\tr(\phi_1^2)$ is singular for $q=-1$, thus
this branch of moduli space does not appear at these roots of unity, and
one only has isolated vacua.

One can also explicitly show that for $q^3 =1$, the element $x_i=
\phi_i^3 + \frac{m^2}{q}\phi_i$ is central. Thus here one gets a large
center, as we have four Casimir operators and one relation. The fourth
Casimir is of the form
\begin{equation}
w = A \phi_1\phi_2\phi_3 + \alpha_1\phi_1^2+\alpha_2\phi_2^2+\alpha_3\phi_3^2
\end{equation}
and it is invariant under the full discrete group of symmetries of the
potential. A Casimir of this form exists for all $q$, and when $m=0$ it is the
familiar $\phi_1\phi_2\phi_3$; it also reduces to the quadratic Casimir
of the $SU(2)$ algebra when $q\to 1$.

The commutative space associated to the algebra is again a deformation
of the $\BC^3/ \BZ_3\times\BZ_3$ orbifold, and it is three complex
dimensional. We have only found a two parameter solution of the
equations; we believe that this is because we chose a very special form
for the solutions, and not necessarily because the ring fails to be
semi-classical.

\subsection{The General Superpotential}\label{sec:general}

In this section we will try to make progress towards understanding the
general deformation, eqs. (\ref{eq:genmarg},\ref{eq:genrelev}). Solving
for the center of the general algebra and also finding the most general
finite dimensional irreducible  representations of the algebra can be
quite difficult. There are some cases which are worth singling out among
these, because at least we can find some partial solutions to  the
moduli space problem. We have also seen that semi-classical rings are
better behaved than others, as they lead to nice commutative geometries.
Finding all possible semi-classical geometries from our sets of
constraints is very important as they might correspond to the behavior
of $D$-branes at new dual singularities (not necessarily orbifolds with
discrete torsion) which can be connected to $AdS_5\times S^5$. Of
particular importance are configurations with conformal invariance, as
they might provide new non-spherical horizons\cite{KW2,MP}. 
Our analysis is quite
incomplete due to the difficulties of the algebraic program involved,
but some general comments will be made here.

As the deformations are taken to zero, the algebra looks like a Poisson
algebra if we interpret commutators as Poisson brackets. Because we have
three variables, and Poisson manifolds are foliated by symplectic 
manifolds (which are of even dimension), the symplectic form in the full
algebra is degenerate and therefore there is at least one constant of
motion. This suggests that there is at least one element of the center
which can be easily computed. For the $q$-deformations, the element of
the center $w = \phi_1\phi_2\phi_3$ exists for arbitrary values of $q$,
which suggests that the element of the center is a polynomial of degree
less than or equal to three, depending on the chosen perturbation.
For marginal deformations, it is indeed of degree three, as will be shown
later; for the deformation by three mass terms it is quadratic
(the Casimir of the $SU(2)$ algebra).

Because we have commutators we can think of the algebra as
deformation-quantization of the Poisson structure. This suggests that we
can standard order operators and establish a correspondence between the
full algebra, and the algebra of three commuting variables. In standard
constructions, this is given by formal power series expansions in a
small parameter $\hbar$. As we have argued before, we want to avoid
infinite power series, and rather give an explicit solution which shows
that the constraints can be standard ordered in some open set. In order
to do this, we separate at each order the polynomials in
$\phi_1, \phi_2, \phi_3$ which can be considered as standard ordered. 

A choice of standard ordering is important. If we want to find elements
of the center, we need to check that their commutators are 
zero for all the generators of the algebra.
Without standard ordering a given expression, it is very hard to decide
if it is zero or not in the algebra.

By
using the constraints, an arbitrary polynomial operator $\CO$ can be
re-ordered into standard ordered form up to small corrections.
We write this as
\begin{equation}
\CO =\CO_{so} + \hbar \CO'
\end{equation}
$\CO_{so}$ is a 
linear combination of standard ordered monomials, and it is
polynomial in $\hbar$. Similarly, we can expand $\CO'\sim a_i M^i$,
where the $a_i$ are polynomial in $\hbar$ and the $M^i$ are a collection
of non-standard ordered monomials.
Because of the form of the algebras, the degree of $\CO'$
as a polynomial in the variables of the algebra is smaller than or equal
to the degree of $\CO$. Taking all the possible non-standard ordered
monomials of degree less than or equal to some fixed number $g$, 
we obtain a matrix equation
\begin{equation}
M^i = M^i_{so} +\hbar a^{i}_j M^j 
\end{equation}
Now $\hbar$ is a small parameter, so the matrix
\begin{equation}
A^j_i = \delta^j_i -\hbar a_i^j 
\end{equation}
is finite dimensional and invertible. Hence, any non-standard
ordered operators may be written
as linear combinations of the standard ordered operators,
where the coefficients are rational functions in the deformation
parameters, the denominators coming from $A^{-1}$.

Since the parameters are complex, more generally 
we need only worry about the possibility
of poles in this construction.
 At such poles, one of two things can happen. Either
the basis for standard ordered polynomials is badly chosen, ({\it e.g.},
the elements become linearly dependent), or there is a true obstruction
to standard ordering independent of the basis. This second  possibility
can happen, if we take $q=0$ for example.

Thus, in principle we can proceed order-by-order in the degree of polynomials
to find central elements. Every element of the algebra can be written in
standard ordered form, and as the degree of the element is preserved or
lowered  by the commutation relations, it is a matter of linear algebra
to calculate the elements of a given order which are in the
center.

Although the procedure is well-defined, it is not efficient, as we need
to calculate the matrix $A$ at each order to resolve this problem. Thus a
general solution of how the center depends on the parameters is at best
difficult to calculate.
Also notice that in the examples we have studied, there is no upper bound
in degree for elements of the center.

In some cases, we may find a large center;
that is, the center is generated by more than one element of the
algebra. If the center is large enough, then we may obtain
a semi-classical algebra.

Let us consider the case where the algebra $\CA$ is a finitely generated
module over its center, with generators $e_i$. We can choose  one of
the generators to be the identity in the ring, and the others will
satisfy a multiplication rule of the type
\begin{equation}\label{eq:eefe}
e_i \cdot e_j =f_{ijk} e_k
\end{equation}
with $f_{ijk}\in \ZA$. On a given irreducible representation of the
algebra, the elements of the center can be treated as numbers, and thus
we can argue that we have a family of algebras parametrized by the
algebraic variety corresponding to the center of the algebra.

Because of the form of eq. (\ref{eq:eefe}), we can see that given
a vector in the representation of the algebra, its orbit under
the action of the $e_i$ is finite dimensional. Thus there is an upper
bound on the dimensions of the irreducible representations. We can
imagine that this upper bound is realized by the branes living in the
bulk, and that any other representation with smaller dimension is a
fractional brane of some sort. The finite dimensionality of the
irreps suggests that the ring is semi-classical, although we have
no proof of this assertion. The semi-classical rings that we have
studied all have this property, and this suggests that the two
conditions might be equivalent.

Let us now consider a few more examples.

\subsubsection{General marginal deformations}

As an example of the general difficulties that one faces, let us consider
a general marginal deformation of the $N=4$ theory.
The superpotential is given by
\begin{equation}
W = \tr\left(\phi_1\phi_2\phi_3 - q \phi_2\phi_1\phi_3 + 
\frac{\lambda}{3} \left(\phi_1^3+\phi_2^3+\phi_3^3\right)\right)
\end{equation}
and the equations we need to solve for the moduli space are (cyclic)
\begin{equation}
\qcomm{\phi_j}{\phi_{j+1}}= -\lambda \phi_{j+2}^2
\end{equation}
This algebra is homogeneous, and thus if we were able to find a
non-trivial irrep, we could scale it to zero: this implies that the
moduli space is connected.

For $\lambda=0$ the element of the center that is always present is $w=
\phi_1\phi_2\phi_3$, and this suggests that the element of the center is
cubic in general. Indeed, a direct calculation shows that
\begin{equation}
(1-q)\phi_1\phi_2\phi_3-\lambda\phi_1^3+q\lambda\phi_2^3-\lambda \phi_3^3
\end{equation}
is central.

Let us first consider one-dimensional irreps. These will satisfy (cyclic)
\begin{equation}
(q-1){\phi_j}{\phi_{j+1}}= \lambda \phi_{j+2}^2
\end{equation}
A non-trivial solution will have the $\phi_j$ all non-zero complex numbers.
We can easily see that this is only solvable provided that
\begin{equation}
(q-1)^3 = \lambda^3
\end{equation}
so $(q-1)/\lambda$ is a cube root of unity. Given $\lambda$ and $q$
satisfying these constraints, one can find solutions to the equations
where $\phi_1$ and $\phi_2$ are equal up to cube roots of unity, and
then $\phi_3$ is determined from the other two. Thus we get three
complex lines meeting at the origin, reminiscent of the moduli space for
$q$-deformations. Indeed when $\lambda$ and $q$ are related in this way,
there is a linear change of basis of the fields which returns the
superpotential to a $q$-deformation. Therefore we have new
semi-classical rings, but they are related by a change of basis to the
ones we already know.

Another thing that we can do is exploit the $\BZ_3$ symmetry which permutes
$\phi_1, \phi_2, \phi_3$. Set $\phi_1 =  a P $, $\phi_2 = b Q$,
$\phi_3 = cP^{-1} Q^{-1}$; for the $P,Q$ matrices associated to the cube
roots of unity, we also have $\phi_3^2 \sim \phi_1\phi_2$, so 
three-dimensional irreps of the
algebra may exist.

In this case we want to find solutions to (cyclic on $a,b,c$)
\begin{equation}
a b (q\omega-1) = \omega \lambda c^2
\end{equation}
with $\omega$ a cube root of unity.
One can see that this gives us the constraint
\begin{equation}
(q\omega-1)^3 = \lambda^3
\end{equation}

For $\lambda\to 0$ we associate the geometry to
$\BC^3/(\BZ_3\times\BZ_3)$ which happens to be one of the orbifolds one
can realize globally on a three-dimensional complex torus.

If for $\lambda\neq 0$ we can find a large center, we might be able to
compute the full geometry of moduli space, and treat this solution as a
fractional brane. Notice that if this is the  case, it does not
correspond to a $\BC^3/\BZ_n\times\BZ_n$ singularity, as the fractional
branes in that case behave differently. Further exploration of this model
will be left for future work\cite{BJL}.

\section{$D$-branes in near-horizon geometries}\label{sec:Dbranes}

So far, we have mainly discussed moduli spaces of vacua and how to
include extended objects in the discussion. The analysis has been done
directly in the field theory. Now  we will try to understand the
background and the moduli space that the $D$-branes realize from the
AdS/CFT perspective.

The field theory is understood as being dual to the near-horizon
geometry of a brane configuration. Because the moduli space is of the
form of a symmetric product (built out of smaller components), one can
think of adding these small component $D$-branes as probes in the near
horizon geometry and testing how the field theory moduli space is
realized on these probes.

We will carefully compare the field theory marginal and relevant
deformations to the corresponding deformations of $AdS_5\times S^5$
geometry. In doing so, we uncover and solve several puzzles. In
particular, there are new branches of moduli space in the field theory
which open up for arbitrarily small values of $q-1$, as discussed in
Section \ref{sec:qdefsec}. Uncovering this structure in the string
theory will have several bonuses. This nongeneric branch is realized by
wrapping a 5-brane on a 2-torus and using this information, we will
argue that the mirror symmetry\cite{BL} between deformed 5-spheres and
orbifolds can be understood as a standard T-duality operation. The two
supergravity descriptions are valid in different areas of parameter space. It
also becomes clear in this analysis that there is no sense in which the
field theories are dual to supergravity on a space; rather, string
theory is absolutely necessary for a consistent duality.

We have seen that the moduli space of vacua has very non-trivial
behavior in response to the deformations. In particular, the somewhat
artificial separation between the center of the algebra and other
elements  of the algebra is very subtle in field theory. This will be
addressed later and we will find a satisfactory solution. If we look at
the same construction from the $AdS_5\times S^5$, each of these
perturbations is in the bulk of the $S^5$ geometry, and there is no
reason to single out any special elements of the algebra.

\subsection{Effects of the Background on $D$-branes}

It is important to notice that as seen from the $AdS_5\times S^5$
perspective when one deforms the theory, the $D$-brane moduli space
changes drastically. To first approximation, this is because the added
potential localizes the $D$-branes to the `fixed planes' of the
deformation. But even for very small deformations $q\sim 1$, we can find
rational solutions of $q^n=1$ for large $n$, and thus the moduli space
has non-generic behavior for a large enough number of branes; indeed, we
need  $n$ such branes to find extra components of moduli space.  These
new branches can be seen from (\ref{eq:decompose}), and predict that the
$D$-branes are going to be uniformly distributed on a circle. We take
the eigenvalues of matrices to determine the coordinates of the
$D$-branes, as in matrix theory\cite{BFSS}.

If the branes are point-like then the open string states stretching
between them would be massive and one would not find the new branch in
moduli space. However, this is clearly inconsistent with our field
theory results, and thus we are motivated to find a satisfactory
solution within string theory. Note that these extra components of
moduli space do not just appear in the vicinity of the origin; 
rather, they extend to
infinity with the rest of $D$-brane moduli space.

The resolution of these issues bears close resemblance to recent results
of Myers\cite{RM} concerning the dielectric properties of branes in
background fields. Since the deformations of the field theory
superpotential correspond to non-zero vevs of fields on the 5-sphere, we
do indeed expect these phenomena to occur. Roughly speaking, the
$D3$-branes should be thought of as $D5$-branes on $\BR^4\times S^2$, where,
as we show below, the $S^2$ is contained in $S^5$. In order to find new
branches of the moduli space, we want to argue that there are
configurations which support massless open string modes, and
topologically this will happen when different spheres intersect each
other. Thus their centers can be separated, and we can still have
massless string states stretching between them.

Now let us begin by  analyzing in some detail the map between
superpotential deformations and vevs. This material is of course not
new, but is included here for completeness.

The $q-1$ and $m$ deformations correspond to background values for
magnetic potentials $F_{(3)}^{RR}$ and $H_{(3)}^{NS}$. The mass
deformation is not marginal, and will therefore depend on the radial
direction of $AdS_5$. The field $\tau=C+ie^{-\phi}$ gives the gauge
coupling, and will be kept constant. The field $G_{(3)} = F_{(3)} -
\langle\tau\rangle H_{(3)}$ is related directly to the superpotential
deformations. The harmonic in the ${\bf 10}$ of $SU(4)$ is a
tachyon state in the $AdS$, thus this perturbation blows-up in the
infrared. The marginal cubic operators correspond to a harmonic of 
$G_{(3)}$ in the ${\bf 45}$ of $SU(4)$. 
In this case, there will be no dependence on the radial
direction of $AdS$ as the associated scalar is massless in five
dimensions; this fact guarantees that we preserve the conformal
group to leading order.

Let us now specialize to the marginal deformations. 
As explained in Ref. \cite{RM}, $D3$-branes in the presence of
$RR$ background fields pick up a dipole moment for
higher brane charge, and become extended in two additional dimensions.
The simplest topological shape, and the one with the lowest energy, is a
2-sphere centered at the position of the $D3$-brane. Since we are
considering a weakly coupled string theory  regime, we should take these to be
$D5$-branes\cite{PS}.  More precisely, the $F_{(3)}$ background is dual
to $F_{(7)}$ which couples to a $D5$-brane. $F_{(7)}$ has support
on $\BR^4\times D^3$, where $D^3$ is the 3-disk with $S^2$ boundary. The
$D3$-branes are extended in the $\BR^4$, which in near-horizon geometry
is contained in $AdS_5$. We thus write $F_{(7)}=\tilde F_{(3)}\wedge
dVol_4$, and integrating, we can normalize it such that
\begin{equation}
\int_{\BR^4\times D^3}F_{(7)}=\int_{D^3}\tilde F_{(3)}
\end{equation}
The 3-disk extends along the radial direction of $AdS_5$ plus two directions
along the  the $S^5$. 
 As a result, we can write
\begin{equation}
\tilde F_{(3)}=d\rho\wedge\tilde C_{(2)}
\end{equation}
As such, if the effect were solely due to the dielectric effect
it is hard to understand how the $D$-branes
can have massless states at different angles along the $S^5$, as the stretching 
happens mostly in the radial direction. 
The $D3$-brane charge of this 5-brane is obtained from a flux through
the 2-sphere, $\frac{1}{2\pi}\int_{S^2}F=n$.

As follows from Ref. \cite{KRN}, there will also be a background
$H_{(3)}^{NS}$ turned on in the presence of the superpotential
deformations. If we expect some energy contribution from the integral of
$H_{(3)}^{NS}$ over the disk, then the 2-sphere prefers to be stretched along
the 5-sphere, because $H_{(3)}^{NS}$ does not have any component along
the $AdS$ directions. In general, then, the radius of the disk $D^3$ is
oriented partially in the radial direction of $AdS_5$ and partially in
$S^5$, as there are two competing effects deforming the branes.

We want to look for configurations where $D3$-branes are intersecting in
the sense of intersections of their $S^2$'s. This is where we can expect
massless string states, at least topologically. The $H^{NS}$ deformation
is the one that gives us the deformation of the $D$-branes in the
appropriate direction. We will assume that these configurations are
supersymmetric and that probes do not affect the background. 

For rational $q$, the moduli space has a scaling direction, which
follows from the conformal invariance: in the language of Section
\ref{sec:qdefsec}, we have
\begin{equation}
a,b,c\to ta,tb,tc
\end{equation}
This is reflected in the near-horizon geometry by the fact that if we
move $D3$-branes along the radial direction of $AdS_5$, they simply
rescale--in particular, if we have intersecting branes, they remain
intersecting as we perform this motion.

For relevant deformations, such as a mass term, the $RR$ and $NS$
backgrounds grow as we move in along the $AdS_5$, and thus we expect the
2-spheres to grow in size along the 5-sphere as we go to the infrared.
As in this case the $H_{(3)}$ fields will also have a radial component, then
both types of fields $H_{(3)}$ and $\tilde F_{(3)}$ 
help the 2-sphere to grow along the $S^5$
and the radial direction. Eventually, the 2-spheres will be of
comparable size to the 5-sphere, and at this point, the notion of
point-like $D$-branes loses any meaning.
To avoid these issues and for ease of calculation, we will treat
only marginal deformations in the following sections.

\subsection{Size and Configurations of 5-branes}

First, let us find the expected size $r$ of an $S^2$ associated with a
$D3$-brane. We will assume that this $S^2$ is small compared to $R_5$,
the radius of $S^5$, but that it is large enough that we can neglect its
self-interaction (from opposite sides of the $S^2$). We will show that
for small deformations, the size grows linearly with the potential. To
this effect, we do a probe calculation. We have a geometry which is
almost $AdS_5\times S^5$ generated by some $D$-branes which are at the
origin, and we have a small extra $D$-brane which turns into a sphere on
which we are going to do our analysis. Because conformal invariance is
preserved by the marginal deformations, there can be no dependence on
the $AdS$ radial direction in the physical quantities of interest, apart
from setting the scale of the physics. We can therefore work in a local
frame and ignore redshift factors, etc.

The DBI action for a $D5$-brane determines the energy 
\begin{equation}\label{eq:D5energ}
E_{DBI}=\frac{\mu_5}{g}\int d^2\Omega\sqrt{\det (G-B+2\pi\alpha' F)}-
\mu_5\int_{D^3}\tilde F_{(3)}-\mu_5\int (2\pi\alpha' F-B)\wedge C_4
\end{equation}
The metric $G$ scales as $r^2$, whereas $F$ behaves as $r^0$ (by the flux
quantization\cite{BDS}). 
By expansion of the DBI part for small $r$, 
we find an energy of the form
\begin{equation}
E=E(D3)+\frac{\alpha}{g} r^4-\beta r^3+o(r^5)
\end{equation}
We write
\begin{equation}
\int_{D_3} H_{(3)}^{NS}=c_{NS} r^3\ \ \ \ \ \ \
\int_{D_3} \tilde F_{(3)}=c_R r^3
\end{equation}
The field strengths are constant over the disk.
We will do the analysis ignoring the five-form field strength. At the
end, we will compensate for this omission. The general features of the
result should not depend on how far we are from the origin. This is how
we can justify this omission.

From the energy (\ref{eq:D5energ}), we see that the $D3$-brane charge
is given by the coupling to $C_4$
\begin{eqnarray}
Q_3&=&n-\frac{1}{4\pi^2\alpha'}\int_{S^2} B\\
&=&n-\frac{c_{NS}}{4\pi^2\alpha'} r^3
\end{eqnarray}

The expansion of the energy in powers of $r$ now reads
\begin{equation}\label{eq:Efull}
E=\frac{\mu_3}{g}\left( Q_3-\frac{c_R}{4\pi^2\alpha'}r^3
+\frac{1}{(4\pi^2\alpha')^2 2n}r^4+\ldots\right)
\end{equation}
where $Q_3$ is a constant plus small coorrections in $r^3$.
The result is minimized at
\begin{equation}
\langle r\rangle\simeq \frac{3}{2}(c_{NS}+gc_R)(4\pi^2\alpha')^2 n
\end{equation}
The energy at this radius satisfies
\begin{equation}\label{eq:energy}
E=\frac{\mu_3}{ g} Q_3 \left(1+
\frac{1}{4Q_3}\langle r\rangle^3(3c_{NS}-gc_R)+\ldots\right)
\end{equation}

This result is puzzling, since it suggests that $n$
$D3$-branes extend to a single $S^2$ of radius proportional to $n$, as
opposed to a sphere wrapped $n$ times around the solution for a single
brane. This
result is wrong from several points of view. First, this solution cannot
give enhanced $U(n)$ gauge symmetry, as there are no massless states
apparent, and suggests a totally different picture of the moduli space, very
different for each value of $n$. We must be more careful in interpreting
eq. (\ref{eq:energy}). 

We interpret the branes as a black hole in the supergravity which is 
almost pointlike as far as the $S^5$ is concerned. One minimizes the
energy (\ref{eq:Efull}) and then compares the ratios of energy to 
$D3$-brane charge of two configurations to determine which may be BPS.
In fact, $n$ $D5$-branes wrapping an $S^2$ of radius $r$
have lower $E/Q_3$ than a single $D5$-brane wrapping an $S^2$ of radius
$nr$. This indicates that the former configuration has the better
chance of being BPS.

The stabilization mechanism which impedes the spheres from shrinking
further is that the flux of $F$ is quantized. This mechanism has been
found when studying $D$-branes from the boundary state formalism for group
manifolds\cite{BDS}, but it is clear that it should be  a general
phenomenon for $D$-branes in non-trivial $H^{NS}_{(3)}$ backgrounds.

Here we see also that the $RR$ charge for the large sphere is not
quantized in general as it gets an anomalous defect proportional to $H^4
n^3$. These can be meta-stable boundary states, and in group manifolds
these can be calculated exactly\cite{BDS}, where a similar defect in
the brane charge quantization condition occurs. When $H_{NS}=0$
the $D$-brane charge is related to K-theory, and then we expect a
quantization condition. This puzzle was recently solved in Ref.
\cite{Taylor} where there is a back reaction from the bulk which 
contributes to the 3-brane charge. Thus, eq. (\ref{eq:energy}) is
incomplete as it does not take into account the energy associated
to this back reaction. However, the ratio $E/Q_3$ on the horizon 
of the brane probe is exact, being the local tension divided by charge. 
Here the BPS $D$-branes behave better, as we get an anomalous charge 
which is proportional to the $D$-brane number.

\subsection{Large $n$ branches}

We now want to find the new branches of moduli space for $q^n=1$, by
finding the geometric configuration into which it can be deformed.
Because the marginal deformation preserves the conformal group, in the
near-horizon geometry, the $H_{(3)}^{NS}$ lies entirely along the
5-sphere, and thus the $D$-brane becomes spherical along 5-sphere
directions.

For the $q$-deformation, this means that the $D$-branes grow in size
linearly with respect to $q-1$, which is the small parameter. This is
the important point of the calculation in the previous section. Consider
a configuration of $n$ of these 2-spheres, distributed around a circle
in $S^5$ so that they touch each other. The value of $n$ is proportional
to $(q-1)^{-1}$, in accordance with $q^n =1$.

In order for the $D$-branes to touch, we need to know the shape of the
2-spheres well. For $|q|=1$, one finds massless states between 2-spheres
which are at the same distance from the origin in $AdS$ space. To see
this, we can calculate the masses of the off-diagonal states from the
superpotential
\begin{equation}
\tr{\phi_1\phi_2\phi_3}- q\tr{\phi_2\phi_1\phi_3}
\end{equation}
with 
\begin{equation}
\phi_1 = \begin{pmatrix} a&0\\
0&b
\end{pmatrix}
\end{equation}
These masses are proportional to $a - qb$ and $b-qa$, and thus in order
to have massless states for $|b|=|a|$ we need $|q|=1$.

In order to get the 2-spheres to touch when they are at different radii,
the dielectric effect on the $D$-brane has to be included, as it is
responsible  for extending the $D$-brane in the radial direction. We now
want to argue that the $D5$-branes laying flat on the $S^5$ actually do
touch at another point. The reason why this is important is that moving
apart a pair of 2-spheres on $S^5$ might make it impossible for them to
touch again. Because of the geometric setup, if we consider two
$D3$-branes at the same location and we move one with respect to the
other in moduli space (of one real dimension on the $S^5$), we will get
two $2$-spheres. Because the solution of the linearized supergravity
equations rotates $H_{(3)}^{NS}$ as we move along this one parameter,
the spheres become linked on the $S^5$. This is explicitly shown in
Figure \ref{fig:knot}.
\FIGURE[4]{\littlefig{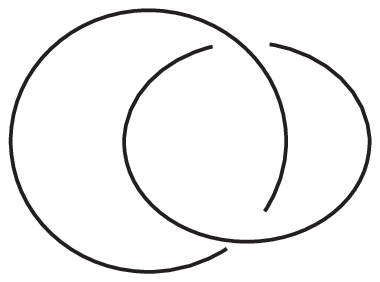}{2}\caption{Knotted spheres on $S^5$}
\label{fig:knot}}
As the spheres are unlinked when they are very far from each other, they
necessarily pass through a point where they touch. This is,
topologically, the place where the extra states become massless. With
the dielectric effect turned on the spheres are tilted with respect to
the $S^5$ and that is why they touch at different values of their radial
position. The tuning required to make the $D$-branes lie flat on the $S^5$
is precisely the action of removing the dielectric effect on the
$D$-branes, and corresponds to one real condition on a one complex
parameter deformation of the theory.

Thus, we arrive at a configuration of spherical $D5$-branes which touch
at points. Now there are configurations, for rational $q$, with $n$
spheres where each touches the next one and they stack on a circle. This
is the configuration where the new branch of moduli space opens up, as
in eq. (\ref{eq:decompose}). This structure should be thought of as a
2-torus with $n$ pinches. Indeed the massless states at the intersection
of the $D$-branes are such that they resolve the pinching points into
tubes, as in Figure \ref{fig:pinching}.
\FIGURE[3]{
\littlefig{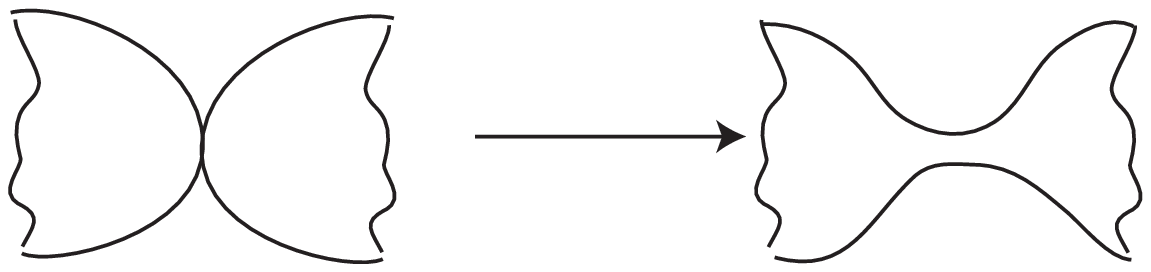}{3}\caption{Resolving the pinched Riemann surface}
\label{fig:pinching}
}
This resolution of these configurations is equivalent to moving onto the
new branches of moduli space.

A pinching torus with $n$ nodes is also exactly the degeneration which
produces fractional branes in an ALE singularity or on an elliptically
fibered Calabi-Yau in F-theory. Thus this configuration of branes seems
to be the right one to deform into the extra branches of moduli space
for the rational values of $q$.

Notice also that this semi-classical torus is reflected also in equations
(\ref{eq:PQ}), where we see a realization of a non-commutative torus
algebra via clock and shift operators.
Thus the non-commutative geometry description of the moduli space knows
that the $D$-brane in $AdS_5\times S^5$ is shaped like a torus, and that
when we deform to the degeneration, we split the torus into $n$ spheres
(fractional branes), as required by the ALE singularity type of the
orbifold in moduli space.

The picture presented above is meant as a topological argument for the
branches of moduli space in string theory. These arguments rely upon a
few technical assumptions, which we think are reasonable. We have
assumed that the different $D$-branes do not affect each other and that
they intersect at supersymmetric angles. Although it would be nice to
assert this, as it would make our whole construction a purely
topological argument, there is no natural complex structure for the
spheres which would guarantee this property, and  we have to rely on a
dynamical mechanism instead. For completeness we should study the
possibility that the 2-spheres might interact strongly with each other
near the intersection point. In that case, the 2-spheres would develop a
throat between them; so, topologically, we have a sphere, with a line
bundle of degree two to count the number of D3 branes. When we move in
moduli space we deform the line bundle and the metric. For a non-generic
bundle, one can get extra states which are massless, and these would be
the extra massless modes one needs. Of course, because the field theory
tells us that the massless states are there, we believe that these
constructions are sensible.

A second point which needs to be made is that although we made an
argument based on $D5$-branes, by the $SL(2,\BZ)$ duality we can make an
argument with any $(p,q)$-5 brane. Thus the fact that the $RR$ and $NS$
mix in the near-horizon geometry is necessary to implement the
$SL(2,\BZ)$ duality on the field theory  space of deformations as we
change the string coupling $g$ and make different $(p,q)$-strings light.
The reason we get a description purely in terms of $D$-branes is that we
are using weakly coupled string theory, and for any other brane with
$NS5$-brane charge the fundamental strings cannot end on it. 
This ambiguity in the description has also been found in Ref. \cite{PS}. 
In their case,
only one configuration would be such that the supergravity degrees
of freedom were weakly coupled through most of the geometry.

\subsection{Mirror Symmetry}

We have seen that the construction of moduli space suggests a two torus 
fibration of the five sphere. This torus can be made explicit by using
the following invariant coordinates
\begin{equation}
r_1^2 = |\phi_1|^2, r_2^2= |\phi_2|^2, r_3^2 = |\phi_3|^2, 
w = \phi_1\phi_2\phi_3
\end{equation}
indeed, $\rho^2=\sum r_i^2$ is the radial direction in $AdS_5$ and $w$ is equal
to $r_1 r_2 r_3$ except for a phase. We get a total of three real
coordinates on the $S^5$, and we are left with two phases to determine,
which are the arguments of $\phi_1/\phi_2$ and $\phi_1/\phi_3$. These
two phases determine the two-torus fibration on the $S^5$, and the
fibration is independent of how many branes are stacked together to
get the new branches of moduli space.

Note that the $T^2$ so obtained may have $n$ nodes (related to the
$D3$-brane charge) but the $T^2$ may wrap $m$ times around the $S^5$
before closing. The latter clearly corresponds to 5-brane charge. 
In Ref. \cite{BL} a mirror symmetry was noted between string theory
on a deformed 5-sphere and an orbifold theory. We are now in position
to demonstrate that this mirror symmetry may be obtained by 
T-duality\footnote{This is expected from the work of Strominger, Yau and
Zaslow\cite{SYZ}.} on the near-horizon geometry, where the 
T-duality is taken fiberwise on the $T^2$ fibration.

The T-duality acts on the 2-torus that we have described above.
Explicitly, the charges $(m,n)$ transform as a doublet under the
$SL(2,\BZ)$ T-duality. Choose a mapping that takes $(m,n)$, with $m,n$
relatively prime, to $(0,1)$; this mapping will single out 
a point-like $D3$-brane on the
mirror. This is achieved by the matrix
\begin{equation}
M=\begin{pmatrix}a&m\\b&n\end{pmatrix}
\end{equation}
where $a,b$ are fixed numbers, modulo $m,n$ respectively. The torus with
complexified K\"ahler form $K=B+iA$ is taken to a torus with a different
value of $K$. Explicitly, we have
\begin{equation}
K\to K'= \frac{aK+m}{bK+n}
\end{equation}

The area of the torus goes to zero and $B_{NS}$ is smooth at the
singularities (where only one phase remains). We can examine the effect
of the transformation on $K$ near this limit. Indeed, we get that in the
dual torus
\begin{equation}
K' \to \frac mn
\end{equation}
which signals a constant $B$-field of strength $m/n$. This value is
quantized and its fractional part corresponds to the discrete torsion
phase.

As the area of the two torus is not constant, if we start without
$H_{NS}$ then upon the T-duality, we will get a varying $B_{NS}$ flux
through the dual torus, and thus we have generated an $H_{NS}$ in the
T-dual geometry. If we want to cancel this quantity, there is a choice
of $B_{NS}$ which makes ${\rm Re}(K') = \frac mn$ constant over the dual
fibration. This determines explicitly the $H_{NS}$ field needed to perform
the marginal deformation on the field theory, from $q=1$ to a given
value of $q$.

Notice that at the singularities we have the allowed degeneration into
fractional branes from the splitting of $(m,n)\to n (m/n, 1)$. Thus the
T-dual fibration has singularities of the $A_{n-1}$ type. As the
fractional branes can be connected to each other in moduli space, we get
a circle of such singularities and the monodromies around that circle
are exactly the ones associated to orbifolds with discrete torsion.

Thus we have both the fractional $B$-field on the T-dual torus, and the
monodromies of the singularities so that we can identify the T-dual
geometry as the orbifold with discrete torsion.

As we have a T-duality description of the relation between the two
compactifications, if the K\"ahler form is generically large in one setup,
it is small in the other. It is therefore necessary to understand which
description can be accounted for by supergravity calculations at a given
point.

This question can be answered in $AdS_5\times S^5 $. If we want $n$
$D$-branes to become one of these 2-tori, then $q^n \sim 1$, and as we
saw before $n\sim 1/ H$. The calculation of the $D$-brane action was done
in string units, thus the $D$-branes are generically of a size
commensurate with the string scale.

When we go to the supergravity regime, the string length is related to the
supergravity background by the relation
\begin{equation}
l_s \sim \frac {1}{\sqrt[4]{g N}}\ l_p.
\end{equation}
In the configurations that we have discussed, we have $nr  \sim R_5\sim\sqrt[4]{gN}$
in string units. Now, in order for $\alpha'$-corrections to be small, we
must have $r\sim H\lesssim \ell_s$, which implies
\begin{equation}
n \gtrsim \sqrt[4]{g N}
\end{equation}
Thus if we want $n$ relatively small, $AdS_5\times S^5$ is
a poor description of the geometry unless the total number of branes
$N$ is such that $\sqrt[4]{gN}<<n$.

Similarly, for the $S^5/\Gamma$ to be large, we need a very large number $N$
of $D$ branes. Indeed, the size of $S^5/\Gamma$ is of order
\begin{equation}
 r \sim l_p/n
\end{equation}
For the supergravity approximation to be valid here, we should require that
twisted sector states are massive; this is the condition $l_s << l_p/n$. As a
result, the crossover region is at the same place, $n\sim\sqrt[4]{gN}$.
Thus, in the orbifold frame, we need to have
$N$ large enough so that $l_p > n l_s$, whereas for the deformed 5-sphere,
 we needed $N$
small enough so that $n l_s > l_p$.

For a general $q$-deformation which is not a root of unity, no
supergravity description will be good, and one is forced to take into
account all of the stringy corrections to the supegravity equations of
motion in order to determine the background.

It is also clear that the strength of the perturbation in string units
needed to change the value of $q$ at the boundary is related to the
number of branes in the configuration. Thus the limit is not uniform in
supergravity. In this sense, it is hard to separate vevs from
expectation values, as the supergravity boundary conditions are changed
drastically when we change the number of branes.

\section{Closed Strings and K-theory}\label{sec:closedstrings}

So far we have described features of the moduli space of vacua for
point-like (in the sense of non-commutative geometry) $D$-branes in
deformed geometries. We want now to present a more complete picture of
the field theory. This will have two aspects. First, we discuss the
chiral ring of the field theory which has a clear interpretation in
terms of the supergravity background and we give an interpretation in
terms of the algebra itself. The second point that we wish to address is
some of the features of extended branes which are accessible by
topological considerations. In particular, this involves a somewhat more
detailed understanding of K-theory and of discrete anomalies.

\subsection{Closed Strings for Near-Horizon Geometry}

Next, we will use ideas from the geometry/field theory correspondence to
describe the physics of closed strings from the field theory point of
view. This closed string theory is to be thought of as the dual string
theory to the field theory of some $D$-branes near a singularity. Our
aim is to understand the open string -- closed string duality a little
better, and how one might expect to realize it in the field theory. We
have dealt with four-dimensional field theories so far in the classical
regime. Our purpose is now to extract a closed string theory out of the
quantum dynamics of the field theory.

The near-horizon geometry will have
certain boundary conditions which control the superpotential, and some
additional set of boundary conditions which specify the vacuum. That is,
there are two contributions to the boundary conditions: those that decay
sufficiently fast are related to the moduli of the branes, and those
that decay more slowly are related to changes in the
superpotential. To fully specify the field theory, we need in addition
the correlation functions of operators. First, though, we need an
identification of those operators.

We take the closed string states to be single trace operators
in the field theory. This is in accordance with the AdS/CFT
correspondence\cite{M, W, GKP} in that closed string states are gauge
invariant operators in the field theory. The idea is to restrict
ourselves now to the chiral ring of the field theory for simplicity, 
and because in all of our analysis we have kept only the parts which are 
protected by supersymmetry.

Let us assume first that we have a conformal field theory, and that its
associated algebra is semi-classical ({\it e.g.}, orbifolds with
discrete torsion). We will exploit the following idea: the vevs of the
closed string states (corresponding to states that decay quickly enough
at the $AdS$ boundary) are generated by the stack of $D$-branes being at
different locations in the moduli space\cite{KW2}. With the asymptotic
values one reconstructs the near-horizon geometry of a set of parallel
$D$-branes by summing over holes\cite{LB2} with given boundary conditions.
Thus we can identify different tadpoles of the string states by motion
in the moduli space of vacua. The right question to ask is what region
of moduli space gives a vev to an operator.

We will combine this knowledge with the identification of the chiral
ring for some geometries. Let us review a few results from Ref. \cite{BL}. In
that paper it was noticed that for orbifolds with discrete torsion, one
could see the twisted and untwisted  string states in the near-horizon
geometry as coming from traces of different chiral operators. We will 
review the case of the orbifold $\BC^3/\BZ_n\times \BZ_n$ with maximal
discrete torsion.

Chiral operators come in two types
\begin{equation}
\CO(k_1,k_2,k_3) = \tr(\phi_1^{k_1}\phi_2^{k_2}\phi_3^{k_3})
\end{equation}
with $k_1 = k_2 = k_3 \mod(n)$, which are untwisted states, and
\begin{equation}
\CO_{j}(k) = \tr\ \phi_{j}^k
\end{equation}
which are twisted states so long as $k\neq 0 \mod(n)$.

The constraint on the $k_j$ for untwisted states is familiar from eq.
(\ref{eq:center}). That is, the center of the algebra is associated with
the untwisted states. This shows why the center of the algebra is so
important to understand the geometry. Namely, the algebraic geometry of
the center of the algebra is the geometry that the closed string sector
sees. Here again we see that the geometry of the closed strings is
commutative, as in Ref. \cite{SW}.
The non-commutativity of the moduli space appears from
the closed string theory point of view because we have twisted sectors.

Notice that in (\ref{eq:fractional}) it is clear that it is the
fractional branes which give vevs to the twisted sector strings. This is
just as it should be, as we always think of coupling twisted sector strings
to fractional branes living at the singularities of the classical space.
Although we have discussed chiral operators here, it is more generally
possible to distinguish twisted and untwisted states. As well, the same
statement may be made if we do not have a conformal field theory.

In the case of $AdS_5\times S^5$, the $F$-term and $D$-term constraints
give us a {\it commutative} geometry. Thus the center is the whole
algebra, and every closed string state is untwisted and
lives in the bulk.

We saw in Section \ref{sec:discrete} that the behavior under mass
deformations was special for $q=\pm1$. From our analysis, we can now
see why this is the case. Namely, for $q=\pm1$, the mass perturbation
is untwisted, and therefore affects the bulk of moduli space. For
any other rational $q$, the mass perturbation is twisted, and we expect
that it will only affect the vicinity of a singularity.

\subsection{Chiral ring revisited and Quantum Groups}

Let us analyze the chiral ring in more detail. We have already learned
that twisted and untwisted states are associated with traces of
central (non-central) elements of the algebra, respectively.

States in the chiral ring are made by taking traces of holomorphic
elements of the algebra.
There are two steps for this construction. First we need a description
of the elements of the algebra, and then we need to interpret the properties of the
trace.

Any operator (for the deformations we have studied) can always be
written in monomial ordered form for a small enough deformation, as we
shown in Section \ref{sec:general}. The difference between two possible
orderings is given by $F$-terms and therefore they correspond to
derivatives of other fields. In a conformal theory, these would be 
descendants and not primaries.  For the topological chiral ring, we set all
$F$-terms to zero, so the operators are identified as traces of 
elements of the algebra.

Let us consider the case where we have a conformal field theory in the
ultraviolet. Because we have an algebra described by quadratic
relations, we have a quantum hyperplane geometry\cite{Manin}. The
operators with the same conformal dimension are homogeneous. On every
quadratic algebra of the type described, there is an associated quantum
group acting on the algebra. The states of the same degree are
associated to the representations of this quantum group. This suggests
that there might be a relation between operators in the closed string
theory and representations of the quantum group. If this is indeed the
case, then the fusion rules of the closed string operators will be
related to the fusion rules of the representations of the quantum group
algebra. This relation would give testable predictions for $3$-point
functions in the deformed $AdS_5\times S^5$ supergravity. Quantum
groups have also made an appearance in near-horizon geometry in the
work of Ref. \cite{JR} in connection with the stringy exclusion
principle\cite{MS}.

We do have to remember that we associate an operator to an element of
the algebra, and that it is not the element of the algebra itself which
is the gauge invariant operator. The association is by taking
\begin{equation}
\CO(a) = \tr(a)
\end{equation}
Because of the cyclic property of the trace, we need the following rule
\begin{equation}
\CO(ab) = \tr(ab) = \tr(ba) = \CO(ba)
\end{equation}
thus the map from the algebra to the operators factors through
\begin{equation}
\CA \to \CA/[\CA,\CA]
\end{equation}
as a vector space. 
It is the class $[a]$ in $\CA/[\CA,\CA]$ that matters, and not $a$ itself.

The space
\begin{equation}
\CA/[\CA,\CA] = HH_0(\CA)
\end{equation}
is actually a homology group of the Hochschild complex\cite{Loday} and
suggests that the chiral ring is in general a cohomology group of the
non-commutative space (so long as we have some sort of Poincar\'e
duality). Because of our knowledge of Calabi-Yau manifolds, we can think
of the chiral ring as a ring of deformations of a non-commutative
complex structure, because we have found a  relation with homology.
Indeed, for a non-compact orbifold space the ring of deformations of
the complex structure is infinite-dimensional because of the
non-compactness, and it is associated to a cohomology group of the
manifold $H^{2,1}(M)$. This suggests that orbifolds with discrete torsion 
may be better understood as a non-commutative Calabi-Yau space.

\subsection{K-theory}

Let us now make a few remarks about K-theory. To this effect we will
review some of the results of Section \ref{sec:discrete}.

Let us analyze the results of the $q$-deformation for rational $q$.
There we found two types of finite dimensional representations:
the representation of a non-commutative point associated to the bulk
and some other representations which correspond to fractional branes at
a singularity.

The set of non-singular points are all connected, and thus each point
defines the same K-theory class. On going to the singularity, the points
would split as
\begin{equation}\label{eq:gensplit}
\lim R_{reg} = \oplus_i R^i_{sing}
\end{equation}
where the subscript indicates that the point belongs to the regular part
of the variety, or the singular part.

It so happens that the $R^i$ are homotopic to each other. That is, they
can be deformed continuously into each other. If $q^n=1$, then there are
$n$ representations on the right hand side of (\ref{eq:gensplit}). In
K-theory we thus have \begin{equation} K(R_{reg}) = n K(R_{sing})
\end{equation} and the K-theory of points is generated by the K-theory
class of a single singular point. Thus $K_0^{p}(\CA) = \BZ$.

If we add one mass deformation, we find two coordinate patches that
cover all of the variety except for a single complex line. This complex
line is the complex line of singularities that was resolved by the
deformation. One can also find solutions that cover this line of singularities.
One still has two complex lines of singularities meeting at the origin,
and the K-theory of points is still $\BZ$. In both of these cases the
degree of a point is enough to determine its K-theory class.

For the other rings, we find different phenomena. There are isolated
points which correspond to fractional branes which cannot be connected
to other singular points. For a rank three mass deformation
and $q=\pm 1$ or $q$ not a root of unity, the moduli space is completely
destroyed and the number of finite dimensional irreducible
representations of the algbera is infinite. These are examples of rings
which are not semi-classical, and in these cases the K-theory of points
consists of an infinite number of copies of $\BZ$, one for each
irreducible representation. In the other cases, for $q$ a root of unity,
the number of isolated points is finite.

The reason why the K-theory of points is not preserved under the
deformations of the algebra relies on the fact that this is the K-theory 
appropriate to algebraic geometry, and not real geometry. This
stems from the fact that we are restricting ourselves to the moduli spce
of vacua, and we are forbidding transitions that go between the
different components in moduli space. This is only appropriate if we are
studying BPS objects, so this K-theory would serve to count BPS
states, and not brane charges.

The full K-theory that we would need to understand brane-charge properly
requires the inclusion of anti-holomorphic data and is less refined.
This new K-theory would be the algebraic K-theory of the $\BC^*$ algebra
associated to the string compactification. That is, the holomorphic
K-theory construction gives too many K-theory classes, and does not give
classes for the objects which cannot be represented in the holomorphic
setting ({\it e.g.}, odd dimensional $D$-branes).

The second statement that we want to make in K-theory has to do with
extended classes. Indeed, based on discrete anomalies, the orbifolds
with discrete torsion have a different K-theory than the commutative one
\cite{W4,W3,FW,Gasp} associated to the ordinary orbifold. Our K-theory
of points reproduces this result. We can also see the anomaly for
extended objects.

Consider the orbifold with discrete torsion $\BC^3/\BZ_n\times \BZ_n$,
and consider trying to wrap a brane along the singular complex line
$x=y=0$. This is a holomorphic subspace of the manifold. For the brane
to cover the complex line, we need to have a lift to the non-commutative
geometry, but the non-commutative geometry covers the singular complex
line by an n-fold cover. Thus if we write a brane solution which would
cover the singular complex line only once (which corresponds to a sheaf
of rank $1$), this solution would correspond to a fractional brane. The
lifting of this solution is obstructed, because if one lifts a point and
does an analytic continuation, the brane would be broken in the
non-commutative space. Indeed, we need a sheaf of rank $1$ in the
non-commutative sense, and this is a sheaf of commutative rank $n$. Thus
the brane charge is quantized in units of $n$ larger than in the
standard orbifold, just as expected from the discrete anomaly. We
believe that the non-commutative analysis makes the calculation of the
anomaly more transparent.

As a final point, note that in principle, we have defined a K-theory
that is capable of extending to $B^{NS}\neq0$.
As seen from the AdS/CFT, the
deformations corresponding to the superpotentials are obtained by
addition of antisymmetric tensors to the background. Our results
suggest that the K-theory necessary to study these background is the
algebraic K-theory of a non-commutative algebra.

\section{Conclusions}\label{sec:conclus}

In this paper, we have studied relevant and marginal deformations of the
$N=4$ SYM theory from a non-commutative algebraic point of view. The
moduli space looks like a symmetric product of a non-commutative
geometry. This is interesting because it implies that $D$-branes may be
considered as independent to a certain extent in the weakly coupled
regime. This symmetric space captures well the phenomena of $D$-brane
fractionation at singularities. Our approach has led us to the
beginnings of a new definition of non-commutative algebraic geometry,
which is still under investigation. The center of the algebra plays an
important role in this construction, and, indeed, in a string theory
picture the commutative subalgebra is related to closed strings, while
the full non-commutative algebra is needed for open strings.
When studied from the AdS/CFT point of view, the field theories that we
studied present new dualities between distinct near-horizon geometries.
These dualities are realized by T-duality of a 2-torus fibration of
the 5-sphere. Different choices of T-duality lead to different dual
near-horizon geometries. These results imply that AdS/CFT is inherently
a stringy phenomenon, as they exhibit T-dualities which are not 
symmetries of classical supergravities. 
In order to understand this duality, we have constructed the $D$-brane
configurations which realize the moduli space. We have found that the
point-like $D3$-branes of the $AdS_5\times S^5$ become non-commutative
5-branes wrapping the torus fibration. 

The non-commutative geometric framework suggests a natural 
formulation of K-theory appropriate to holomorphic data, and this is
successful in reproducing the physics of discrete anomalies. This 
suggests that in general backgrounds the K-theory appropriate to 
$D$-brane charge is that derived from non-commutative algebra.

Our work suggests several avenues for future research. In particular,
it would be of interest to understand the general problem of classifying
what we have termed semi-classical algebras. A thorough understanding
of this problem should provide new backgrounds in which $D$-branes
can propagate, and would shed light on the existence of other
dualities in near-horizon geometries. 

In more generality, one should study the full problem of non-commutative
algebraic geometry, including global questions. With a precise notion of
gluing and compactness for example, we could entertain the idea of
non-commutative Calabi-Yaus and their stringy geometry.

There are also interesting questions concerning non-perturbative 
effects, which we would need to understand $S$-dualities for example.
We also must be concerned about the possibility of non-perturbative
effects modifying our results, through, for example, the appearance
of multi-trace operators in the superpotential. 

For relevant deformations, there will be renormalization group flows
which are reflected in near-horizon geometry. It would be of interest 
to construct these flows for the examples that we have studied, particularly
since one expects that stringy corrections become important in the 
infrared. The study of correlation functions should also be of interest,
with a possible connection to the representation theory of quantum groups.

Generalizations of our work to more complicated quivers is possible and
will be explored elsewhere. 

\bigskip
\noindent {\bf Acknowledgments:} We wish to thank M. Ando, D. Grayson,
A. Hashimoto, A. Jevicki, D. Kutasov, J. Maldacena, A. Strominger and C.
Vafa for discussions. We are especially indebted to M. Strassler for a
critical reading of this paper. DB thanks Harvard University for
hospitality. Work supported in part by U.S. Department of Energy, grant
DE-FG02-91ER40677 and an Outstanding Junior Investigator Award.

\providecommand{\href}[2]{#2}\begingroup\raggedright\endgroup

\end{document}